\documentclass[conference, a4paper]{IEEEtran}
\IEEEoverridecommandlockouts
\usepackage{cite}
\usepackage{amsmath,amssymb,amsfonts}
\usepackage{graphicx}
\usepackage{textcomp}
\usepackage{xcolor}
\def\BibTeX{{\rm B\kern-.05em{\sc i\kern-.025em b}\kern-.08em
    T\kern-.1667em\lower.7ex\hbox{E}\kern-.125emX}}
    
\usepackage{subcaption,setspace,booktabs,epstopdf,dblfloatfix,blindtext,amsthm}
\usepackage{algorithm,algcompatible,upgreek,acro}
\usepackage[binary-units=true]{siunitx}
\usepackage{multirow}

\usepackage{amsmath,amssymb}
\newcommand{\bbR}{{\mathbb{R}}}

\newcommand{\bs}{{\mathbf{s}}}

\newcommand{\bx}{{\mathbf{x}}}

\newcommand{\bR}{{\mathbf{R}}}

\newcommand{\bX}{{\mathbf{X}}}

\newcommand{\rmm}{{\mathrm{m}}}

\newcommand{\rmB}{{\mathrm{B}}}

\newcommand{\rmE}{{\mathrm{E}}}

\newcommand{\rmN}{{\mathrm{N}}}

\newcommand{\rmS}{{\mathrm{S}}}
\newcommand{\rmT}{{\mathrm{T}}}
\newcommand{\rmU}{{\mathrm{U}}}

\newcommand{\rmW}{{\mathrm{W}}}

\newcommand{\cN}{\mathcal{N}}

\newcommand{\cU}{\mathcal{U}}

\newcommand{\bmu}{\boldsymbol{\mu}}

\newcommand{\bSigma}{\boldsymbol{\Sigma}}

\newcommand{\transp}{{\sf T}}

\makeatletter
\def\munderbar#1{\underline{\sbox\tw@{$#1$}\dp\tw@\z@\box\tw@}}
\makeatother
\newcommand{\dBm}{\SI{}{\decibel}\rmm}

\DeclareAcronym{3GPP}{
  short=3GPP,
  long=3rd generation partnership project
}
\DeclareAcronym{ADC}{
  short=ADC,
  long=analog-to-digital converter
}
\DeclareAcronym{AMP}{
  short=AMP,
  long=approximate message passing
}
\DeclareAcronym{ANN}{
  short=ANN,
  long=artificial neural network
}
\DeclareAcronym{AoA}{
  short=AoA,
  long=angle-of-arrival
}
\DeclareAcronym{AoD}{
  short=AoD,
  long=angle-of-departure
}
\DeclareAcronym{APS}{
  short=APS,
  long=azimuth power spectrum
}
\DeclareAcronym{AR}{
  short=AR,
  long=augmented reality
}
\DeclareAcronym{AV}{
  short=AV,
  long=autonomous vehicle
}
\DeclareAcronym{BM}{
  short=BM,
  long=beam management
}
\DeclareAcronym{BS}{
  short=BS,
  long=base station
}
\DeclareAcronym{BSM}{
  short=BSM,
  long=basic safety message
}
\DeclareAcronym{BW}{
  short=BW,
  long=bandwidth
}
\DeclareAcronym{CDF}{
  short=CDF,
  long=cumulative distribution function
}
\DeclareAcronym{CP}{
  short=CP,
  long=cyclic-prefix
}
\DeclareAcronym{CSI-RS}{
  short=CSI-RS,
  long=channel state information reference signal
}
\DeclareAcronym{DFT}{
  short=DFT,
  long=discrete Fourier transform
}
\DeclareAcronym{DL}{
  short=DL,
  long=downlink
}
\DeclareAcronym{EKF}{
  short=EKF,
  long=extended Kalman filter
}
\DeclareAcronym{DSRC}{
  short=DSRC,
  long=dedicated short-range communication
}
\DeclareAcronym{FDD}{
  short=FDD,
  long=frequency division duplex
}
\DeclareAcronym{FMCW}{
  short=FMCW,
  long=frequency modulated continuous wave
}
\DeclareAcronym{FoV}{
  short=FoV,
  long=field-of-view
}
\DeclareAcronym{GNSS}{
  short=GNSS,
  long=global navigation satellite system
}
\DeclareAcronym{IMU}{
  short=IMU,
  long=inertial measurement unit
}
\DeclareAcronym{lidar}{
  short=lidar,
  long=light detection and ranging
}
\DeclareAcronym{LOS}{
  short=LOS,
  long=line-of-sight
}
\DeclareAcronym{LPF}{
  short=LPF,
  long=low pass filter
}
\DeclareAcronym{LTE}{
  short=LTE,
  long=long term evolution
}
\DeclareAcronym{MIMO}{
  short=MIMO,
  long=multiple-input multiple-output
}
\DeclareAcronym{ML}{
  short=ML,
  long=machine learning
}
\DeclareAcronym{mmWave}{
  short=mmWave,
  long=millimeter wave
}
\DeclareAcronym{MRR}{
  short=MRR,
  long=medium range radar
}
\DeclareAcronym{NLOS}{
  short=NLOS,
  long=non-line-of-sight
}
\DeclareAcronym{NB}{
  short=NB,
  long=narrow beam
}
\DeclareAcronym{NR}{
  short=NR,
  long=new radio
}
\DeclareAcronym{OFDM}{
  short=OFDM,
  long=orthogonal frequency-division multiplexing
}
\DeclareAcronym{ppm}{
  short=ppm,
  long=parts-per-million
}
\DeclareAcronym{PF}{
  short=PF,
  long=particle filter
}
\DeclareAcronym{RMS}{
  short=RMS,
  long=root-mean-square
}
\DeclareAcronym{RPE}{
  short=RPE,
  long=relative precoding efficiency
}
\DeclareAcronym{RS}{
  short=RS,
  long=reference signal
}
\DeclareAcronym{RSRP}{
  short=RSRP,
  long=reference signal received power
}
\DeclareAcronym{RSU}{
  short=RSU,
  long=roadside unit
}
\DeclareAcronym{SCS}{
  short=SCS,
  long=subcarrier spacing
}
\DeclareAcronym{SNR}{
  short=SNR,
  long=signal-to-noise ratio
}
\DeclareAcronym{SSB}{
  short=SSB,
  long=synchronization signal block
}
\DeclareAcronym{THz}{
  short=THz,
  long=terahertz
}
\DeclareAcronym{UAV}{
  short=UAV,
  long=unmanned aerial vehicle
}
\DeclareAcronym{UE}{
  short=UE,
  long=user equipment
}
\DeclareAcronym{UKF}{
  short=UKF,
  long=unscented Kalman filter
}
\DeclareAcronym{UL}{
  short=UL,
  long=uplink
}
\DeclareAcronym{ULA}{
  short=ULA,
  long=uniform linear array
}
\DeclareAcronym{V2I}{
  short=V2I,
  long=vehicle-to-infrastructure
}
\DeclareAcronym{V2V}{
  short=V2V,
  long=vehicle-to-vehicle
}
\DeclareAcronym{V2X}{
  short=V2X,
  long=vehicle-to-everything
}
\DeclareAcronym{VR}{
  short=VR,
  long=virtual reality
}
\DeclareAcronym{VRU}{
  short=VRU,
  long=vulnerable road user
}
\DeclareAcronym{WB}{
  short=WB,
  long=wide beam
}
\DeclareAcronym{RNN}{
	short=RNN,
	long=recurrent neural network
}
\DeclareAcronym{LSTM}{
	short=LSTM,
	long=long short-term memory
}

\DeclareAcronym{FC}{
	short=FC,
	long=fully connected
}
    
\algnewcommand\INPUT{\item[\textbf{Input:}]}
\algnewcommand\OUTPUT{\item[\textbf{Output:}]}
\algnewcommand\INIT{\item[\textbf{Initialization:}]}

\newcommand{\MBS}{M_{\rmB\rmS}}
\newcommand{\MUE}{M_{\rmU\rmE}}

\newcommand{\TSS}{T_{\rmS\rmS}}
\newcommand{\PT}{P_{\rmT}}
\newcommand{\powtodb}{{\mathrm{pow2db}}}
\newcommand{\dbtopow}{{\mathrm{db2pow}}}
\DeclareSIUnit \dB {\decibel}
\DeclareSIUnit \dBm {dBm}

\begin{document}

\title{Beam Management with Orientation and RSRP using Deep Learning for Beyond 5G Systems}

\author{\IEEEauthorblockN{Khuong N. Nguyen,
Anum Ali, Jianhua Mo, Boon Loong Ng, Vutha Va, and Jianzhong Charlie Zhang}
\IEEEauthorblockA{Standards and Mobility Innovation Laboratory, Samsung Research America, Plano, TX 75023 USA\\
% Email: k.nguyen1, anum.ali, jianhua.m, b.ng, vutha.va, jianzhong.z@samsung.com}}
Email: \{k.nguyen1,~anum.ali,~jianhua.m,~b.ng,~vutha.va,~jianzhong.z\}@samsung.com}}

% \author{\IEEEauthorblockN{1\textsuperscript{st} Khuong N. Nguyen}
% \and
% \and
% \and
% \and
% \and

\maketitle

\begin{abstract}
Beam management (BM), i.e., the process of finding and maintaining a suitable transmit and receive beam pair, can be challenging, particularly in highly dynamic scenarios. Side-information, e.g., orientation, from on-board sensors can assist the user equipment (UE) BM. In this work, we use the orientation information coming from the inertial measurement unit (IMU) for effective BM. We use a data-driven strategy that fuses the reference signal received power (RSRP) with orientation information using a recurrent neural network (RNN). Simulation results show that the proposed strategy performs much better than the conventional BM and an orientation-assisted BM strategy that utilizes particle filter in another study. Specifically, the proposed data-driven strategy improves the beam-prediction accuracy up to $34\%$ and increases mean RSRP by up to $\SI{4.2}{dB}$ when the UE orientation changes quickly.
\end{abstract}


\section{Introduction}
\label{sec:introduction}
%%%%%%%%%%%%%%%%%%%%%%%%%%%%%%%%%%%%%%%%%%%%%%%%%%%%%%%%%%%%%%%%%%%%%%%%%%%%%%%%%%%%%%%%%%%%%%%%%%%%%%%%%%%%%
%Par 1: General introduction of the area plus what we do
Communication at~\ac{mmWave} and \ac{THz} frequencies is suitable for high data-rate applications due to the availability of large bandwidth~\cite{rappaport2013millimeter,giordani2020toward}. The use of large antenna arrays with beamforming at the transmitter and receiver is needed to achieve an adequate link margin and to overcome the high free-space path loss \cite{Bjornson_MWC19}. The process of identifying and maintaining a suitable beam pair for the link is known as \ac{BM} ~\cite{Giordani2019Tutorial, Li_Access20}. Successful \ac{BM} is difficult, particularly in the urban area and highly mobile scenarios where the channel changes frequently \cite{Heng_COMM21}. 

Due to the strong fitting ability, \ac{ML} has been adopted as a promising solution for \ac{mmWave} beam alignment, which is inherently a complex nonlinear problem. The context information, such as \ac{UE} locations \cite{Va_TVT18, Maschietti_GC17, Heng_TCCN21}, sub-6 GHz out-of-band information \cite{Ali2018Millimeter, Alrabeiah_TCOM20}, could be used to predict the best beam. 
By capturing the temporal correlation of mmWave channels experienced by the mobile \ac{UE}, the \ac{LSTM} based methods were proposed for beam tracking and proactive beam switching in \cite{kaya2021deep, Lim_TCOM21, Echigo_TVT21, Ma_Ke_TCOM21}. 
%\cite{Heng_TCCN21} proposed to predict the BS and beam given the UE position.
%\cite{Ma_Wenyan_WCL20} %\cite{Heng_TWC22} 
These work, however, did not consider the 3D orientation of a hand-held UE which could rotate quickly in daily usage, e.g., from the portrait to the landscape mode.

% Par 2: Contributions that we have
In this paper, we use onboard sensor information as additional information to improve performance and achieve efficient \ac{BM} at the \ac{UE}. Specifically, we jointly use the orientation information from the \ac{IMU} and the \ac{RSRP} for \ac{BM}. The orientation information is readily available since \acp{IMU} are used in most consumer electronic devices, such as mobile phones, \ac{AR}/\ac{VR} gadgets, and \acp{UAV}.
The main contribution of this work is that we developed a method to utilize both \ac{RSRP} and orientation information to achieve successful \ac{BM} at the \ac{UE} with the help of deep learning. The orientation-assisted \ac{BM} problem is formulated as a classification problem, with \ac{RSRP} and orientation information as inputs and beam index as output. A \ac{RNN} is used for classification. The proposed formulation is consistent with $5$G \ac{NR} signaling and does not require any modifications to the standard. We present simulation results using a practical \ac{UE} beam codebook and realistic ray-tracing multi-path channels. These practical assumptions make the evaluation results more credible. Compared to the case without orientation information, the proposed strategy improves beam prediction accuracy by up to $34 \%$ and mean \ac{RSRP} by up to $\SI{4.2}{dB}$ when the UE rotates quickly.

% Part 4: paper organization % JMo: not needed for a conference paper
The remainder of this paper is organized in the following manner. The related work is discussed in the next section. The communication system model, the \ac{RSRP}, and the orientation model are discussed in Section \ref{sec:system_model}. In Section \ref{sec:simulation_setup}, we present the simulation setup that we used to generate the data to train the \ac{RNN} model. The proposed deep learning based BM strategy is described in detail in Section \ref{sec:dt_beam_management}. We present numerical results in Section \ref{sec:simulation_results} to demonstrate the effectiveness of the proposed strategy. Finally, Section \ref{sec:conclusion} concludes the paper.

\section{Related Work}
\label{sec:related_work}
%Par 3: Literature review: Only IMU assisted work - our pf based work will also become prior work
There has been some previous work on \ac{BM}~\cite{shim2014application,qi2018three,brambilla2019inertial,Rezaie2020Location,ali2021orientation} using orientation information. Given the best beam before orientation change and the orientation change, the best beam after orientation change is predicted in~\cite{shim2014application}. Specifically, first, the \ac{AoA} is obtained from the best beam, and then to predict the best beam after orientation change, the change in orientation is converted into a change in \ac{AoA}. The orientation and position of the \ac{UE} relative to the \ac{BS} are tracked and used for beam steering in~\cite{qi2018three}. The change in pitch is used to maintain the \ac{LOS} link between two vehicles in ~\cite{brambilla2019inertial}. Location and orientation are jointly used in an \ac{ML}-based inverse fingerprinting method for \ac{BM} in~\cite{Rezaie2020Location}. Finally,~\cite{ali2021orientation} employs a \ac{PF} to combine the orientation and \ac{RSRP} information for \ac{BM}.

%Par 4: Differences: If can't differentiate from ML work mention codebook and signalling
The earlier work on using orientation information for \ac{BM} has several shortcomings. In particular, the strategy of~\cite{shim2014application} can work only if the \ac{AoA} aligns with the best beam's peak, which is not guaranteed. As a result, any prediction based on \ac{AoA} that is incorrect is also likely to be sub-optimal. The beam steering method of~\cite{qi2018three}, i.e. relative position/orientation tracking, is useful only in \ac{LOS}. Furthermore, the beam steering ignores the hardware limitations of current \ac{mmWave} systems. Only in vehicular context, where the leading and following vehicles have a strong \ac{LOS} path and change in pitch is the primary source of change in the \ac{LOS} path, is the strategy of \cite{brambilla2019inertial} useful. Unlike this work, the techniques of ~\cite{shim2014application,qi2018three,brambilla2019inertial,Rezaie2020Location} do not take into account the $5$G \ac{NR} signaling and realistic beam codebooks. 

%Finally, the same setup as in this study is used in the prior work ~\cite{ali2021orientation}. In comparison to ~\cite{ali2021orientation} that uses particle filter ~\cite{kunsch2013particle}, however, this proposed strategy uses a data-driven approach that employs deep learning, and simulation results show that the proposed data-driven approach performs better than the \ac{PF}-based strategy used in ~\cite{ali2021orientation}.

Finally, the same setup as in this study is used in our prior work \cite{ali2021orientation}. In comparison to \cite{ali2021orientation} that uses particle filter \cite{kunsch2013particle}, however, this proposed strategy uses a data-driven approach that employs deep learning. Although the particle filter algorithm has some advantages such as its high stability with tracking problems and its ability to implicitly track the gain which eliminates the need to have a process model for the gain evolution, it also has its shortcomings. Particularly, its non-deterministic characteristic might lead to wrong predictions when uninformative sensor readings are collected for an extended period. Additionally, its computational intensive requirement (a good filter requires a significant amount of particles) will put quite a burden on the \ac{UE}. On the contrary, using deep learning puts a less computational load on the \ac{UE} and its end-to-end characteristic eliminates the need to do feature engineering and makes the classifier system simpler. More importantly, the experimental results in this study confirm that the proposed data-driven approach performs much better than the \ac{PF}-based strategy used in \cite{ali2021orientation}.

%Par 5: Notation
\emph{Notation}: For column vectors, bold lowercase $\bx$ is used, bold uppercase $\bX$ is used for matrices, and non-bold letters $x, X$ are used for scalars. $[\bx]_j$ is the $j$th entry in a column vector $\bx$. The transpose and conjugate transpose are represented by the superscripts $\transp$ and $\ast$, respectively. %$\bI_N$ represents the $N\times N$ identity matrix.
$\cN(\bmu,\bSigma)$ is a multidimensional Gaussian random variable with mean $\bmu$ and covariance $\bSigma$. $\cU[a,b]$ is a Uniform random variable with support $[a,b]$. The modulo operator is denoted by the symbol $\mod(\cdot)$. We define $\powtodb(x)=10\log_{10}(x)$ and $\dbtopow(x)=10^{\frac{x}{10}}$ functions to transform powers from linear to logarithmic scale and back.
%%%%%%%%%%%%%%%%%%%%%%%%%%%%%%%%%%%%%%%%%%%%%%%%%%%%%%%%%%%%%%%%%%%%%%%%%%%%%%%%%%%%%%%%%%%%%%%%%%%%%%%%%%%%%

\section{System model}
\label{sec:system_model}
\begin{figure}[t]
	\centering
    \includegraphics[width=0.47\textwidth]{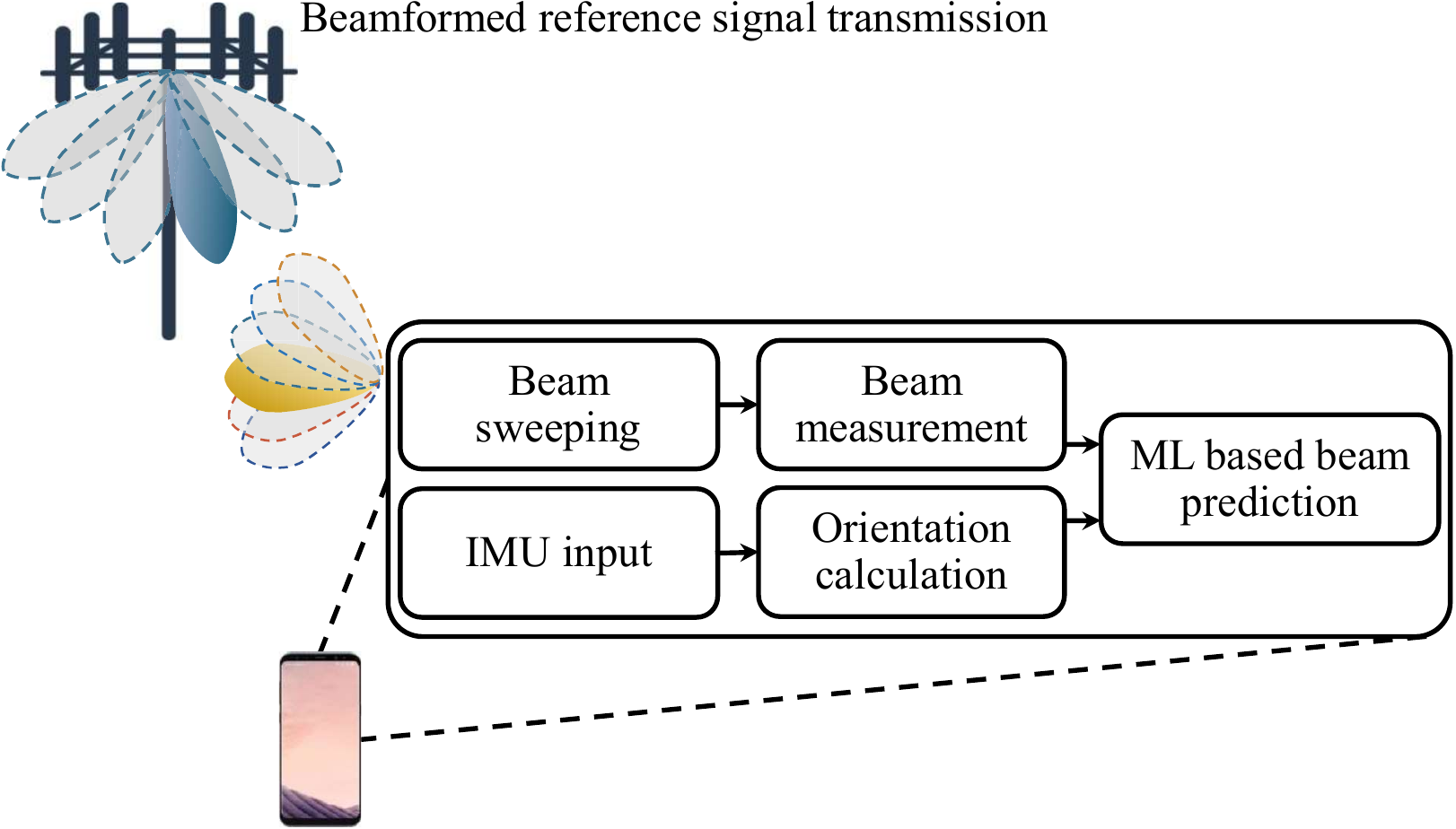} % for arxiv
	\caption{The block diagram of a \ac{BS}-\ac{UE} communication system in which the \ac{UE} uses beam measurements as well as orientation information for beam prediction.}
	\label{fig:BD}
\end{figure}

%%%%%%%%%%%%%%%%%%%%%%%%%%%%%%%%%%%%%%%%%%%%%%%%%%%%%%%%%%%%%%%%%%%%%%%%%%%%%%%%%%%%%%%%%%%%%%%%%%%%%%%%%%%%%
% Par 1: General system model
We consider a communication system shown in Fig.~\ref{fig:BD} where the \ac{RSRP} information, i.e., information extracted from beam measurements, as well as orientation information, is used at the \ac{UE} side for beam prediction.

% Par 2: RSRP information - no need to make subsections plus reduce the text
%--------------------------------------------------------------------------------------------------------------
\subsection{\ac{RSRP} information}\label{sec:RSRPinfo}
%--------------------------------------------------------------------------------------------------------------

\begin{figure}[t]
	\centering
	\begin{subfigure}{0.23\textwidth}
		\centering
		\includegraphics[width=1\textwidth]{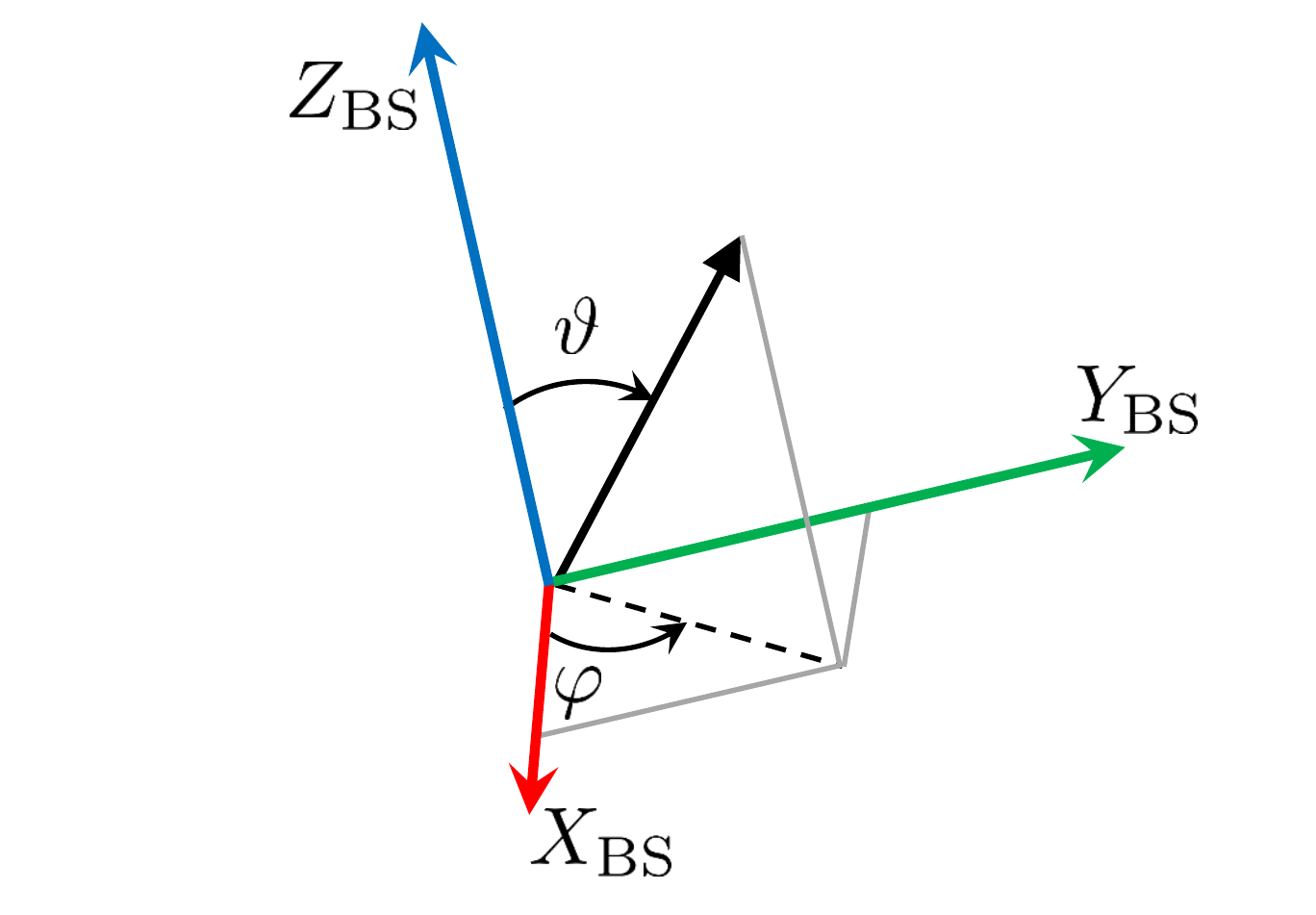}
		\caption{\ac{BS} local coordinate system.}
		\label{fig:Coordinate_system_BS}
	\end{subfigure}
	\begin{subfigure}{0.23\textwidth}
		\centering
		\includegraphics[width=1\textwidth]{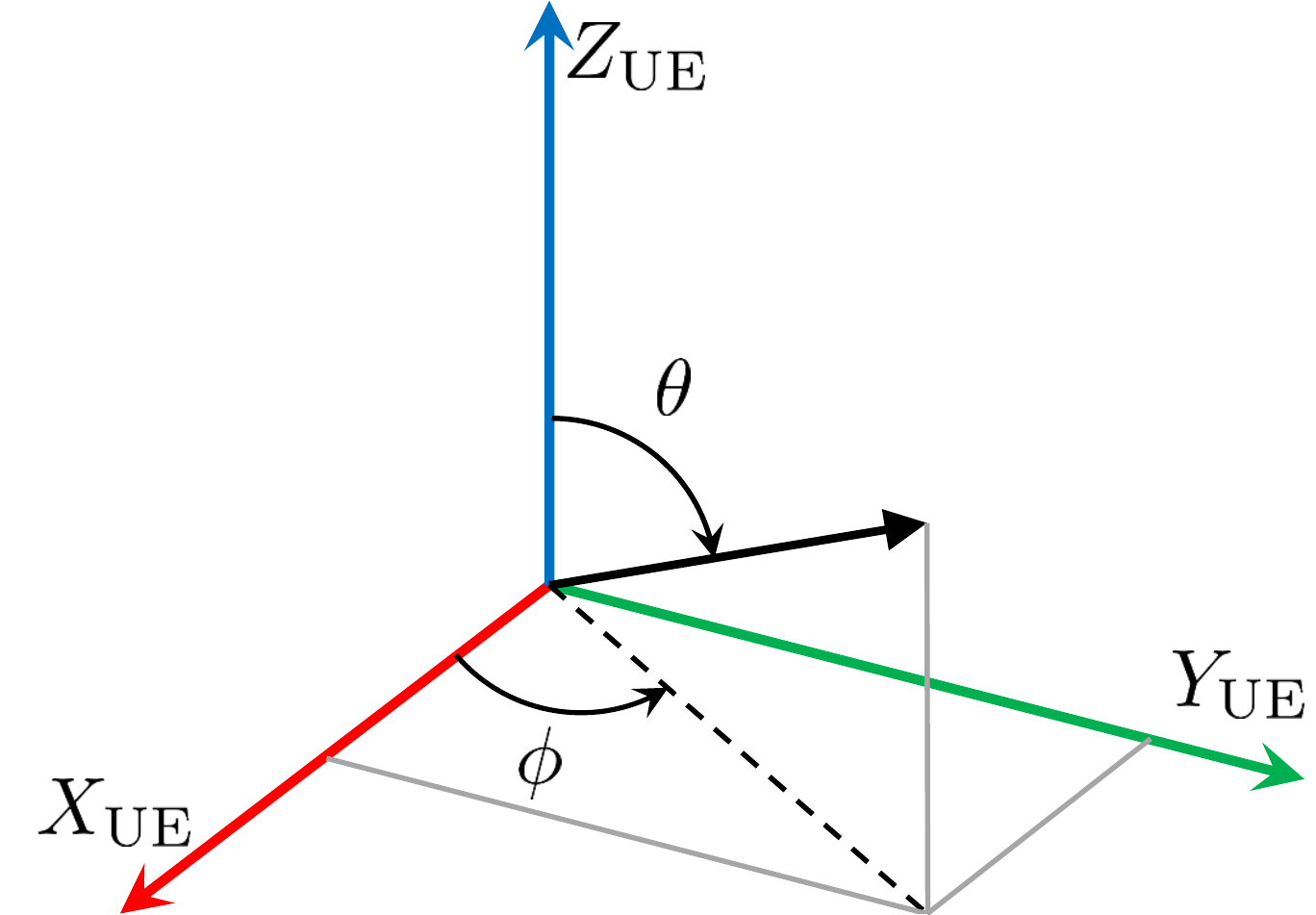}
		\caption{\ac{UE} local coordinate system.}
		\label{fig:Coordinate_system_UE}
	\end{subfigure}
	\caption{Local coordinate systems of the \ac{BS} and \ac{UE}.}
	\label{fig:Coordinate_systems}
\end{figure}

For \ac{DL} \ac{BM} in $5$G \ac{NR}, the \ac{BS} sends the beam-formed \acp{SSB} and/or \acp{CSI-RS}. The \ac{UE} receives the beam-formed transmissions using different receive beams. The \ac{UE} then reports the index and quality of the best beams to the \ac{BS}. Subsequently, the \ac{BS} decides the best transmit beam. In this work, we consider \ac{SSB} based \ac{DL} \ac{BM}. The transmission power is $\PT~\SI{}{\dBm}$, and the coordinate systems are shown in Fig.~\ref{fig:Coordinate_systems}. Specifically, the local coordinate system of the \ac{BS} is shown in Fig.~\ref{fig:Coordinate_system_BS}. The transmit beam codebook contains $\MBS$ beams. If the \ac{BS} uses transmit beam $i\in{1,\cdots,\MBS}$, then the transmit beam gain in direction $(\varphi,\vartheta)$ is $F_i(\varphi,\vartheta)$ $\SI{}{\dB}$, where $\varphi$ is the azimuth angle and $\vartheta$ is the zenith angle. Similarly, the receive beam code-book contains $\MUE$ code-words (or beams), and the local coordinate system used at the \ac{UE} is shown in Fig.~\ref{fig:Coordinate_system_UE}. If the \ac{UE} uses receive beam $j\in{1,\cdots,\MUE}$, then the receive beam gain in direction $(\phi,\theta)$ is  $G_j(\phi,\theta)$ $\SI{}{\dB}$, where $\phi$ is the azimuth angle and $\theta$ is the zenith angle. 
	
The \acp{SSB} are transmitted with periodicity $\TSS$, and the time variable $t$ denotes the index of an \ac{SSB}. The received \ac{RSRP} at time $t$, i.e., $s_t$ is given as
%
%\scalebox{0.95}{$
\begin{align}
s_t = & \PT + \powtodb \left(\frac{\text{SCS}}{\text{BW}} \right) + \powtodb \Big(\sum_{c=1}^C  \dbtopow \big(p_t^{(c)} + \nonumber \\ 
& \quad F_{i_t}(\varphi_t^{(c)},\vartheta_t^{(c)}) +  G_{j_t}(\phi_t^{(c)},\theta_t^{(c)})\big) \Big) + n_t %\nonumber
\end{align}
%$}
where \ac{SCS} is for the \ac{SSB}, and \ac{BW} is for the system, since the \ac{RSRP} is defined as the average received power over a single sub-carrier. Further, $C$ is the number of multi-paths in the channel, and $p_t^{(c)}$ $\SI{}{\dB}$ is the path gain of the $c$-th path at time $t$. The \ac{AoD} of the $c$th path is $(\varphi_t^{(c)},\vartheta_t^{(c)})$ and the \ac{AoA} of the $c$th path is $(\phi_t^{(c)},\theta_t^{(c)})$. Finally, $i_t$ is the index of the transmit beam at time $t$, $j_t$ is the index of the receive beam at time $t$, $F(\phi, \theta)$ is the BS beam pattern, $G(\phi, \theta)$ is the UE beam pattern, and $n_t$ is the noise.	
	
In this work, we assume the genie-aided knowledge of best transmit beam, i.e., ${i^\star}$ and focus on finding ${j^\star}$, as our main focus is on the use of \ac{UE} orientation for \ac{BM}. Under this assumption, the conventional \ac{RSRP}-only based \ac{BM} works in the following manner. An \ac{RSRP} table $\bs\in\bbR^{\MUE\times 1}$ (with all entries initialized to $-\infty$) is maintained at the UE. The \ac{UE} receives the \ac{SSB} through different beams in a round-robin manner. At time $t$, the $j_t$-th beam is used as the receive beam, and the $j_t$-th entry of the \ac{RSRP} table is subsequently updated as $[\bs]_{j_t}=s_t \label{eq:RSRPtable}$.

The best receive beam $\hat j^\star$ can then be bound as $\hat j^{\star}=\underset{j}{\arg\max}~[\bs]_j$. As it takes $\MUE$ \ac{SSB} periods to receive updated \ac{RSRP} information on all beams, some entries in the \ac{RSRP} table may be outdated due to temporal changes, e.g., \ac{UE} orientation.	

%--------------------------------------------------------------------------------------------------------------
\subsection{Orientation information}\label{sec:IMUinfo}
\begin{figure}[t!]  % JMo: This figure is not important. Skip it to save the space.
	\centering
	\includegraphics[width=0.35\textwidth]{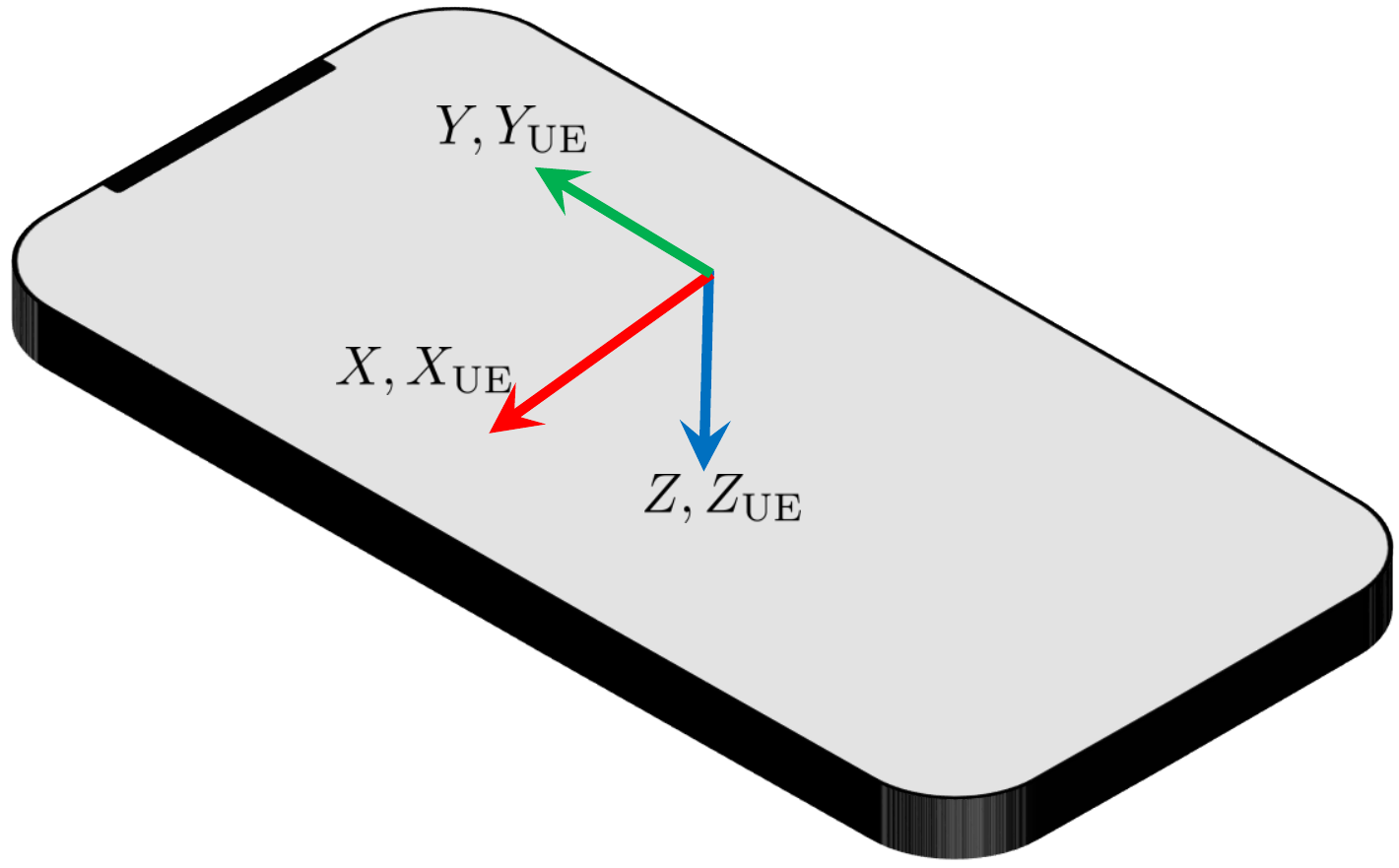}
	\caption{The orientation of the phone when the \ac{UE} local coordinate system $X_{\rmU\rmE},Y_{\rmU\rmE},Z_{\rmU\rmE}$ is aligned with global coordinate system $X,Y,Z$.}
	\label{fig:Phone}
\end{figure}

%--------------------------------------------------------------------------------------------------------------
Our objective is to use orientation information coming from an \ac{IMU} together with the \ac{RSRP} information for \ac{BM} as in Fig.~\ref{fig:BD}. 
The local coordinate system ($X_{\rmU\rmE},Y_{\rmU\rmE},Z_{\rmU\rmE}$), and the global coordinate system ($X,Y,Z$) of the \ac{UE} are shown in Fig.~\ref{fig:Phone}, which are aligned. 
The rotations around $Z$, $Y$, and $X$ are denoted by $\alpha$, $\beta$, and $\gamma$ respectively~\cite{3GPP38901}, and the rotations are assumed in that order~\cite[Section 7.1]{3GPP38901}. The rotation matrices around each  axis, i.e., $\bR_Z(\alpha)$, $\bR_Y(\beta)$, and $\bR_X(\gamma)$ are defined as

% \begin{align}
% \scalebox{0.9}{
% $
% 	\bR_Z(\alpha)=
% 	\begin{bmatrix}
% 		\cos\alpha  & -\sin \alpha & 0 \\
% 		\sin \alpha & \cos\alpha   & 0 \\
% 		0           & 0            & 1
% 	\end{bmatrix};
% 		\bR_Y(\beta)=
% 	\begin{bmatrix}
% 		\cos\beta  & 0 & \sin \beta \\
% 		0          & 1 & 0          \\
% 		-\sin\beta & 0 & \cos\beta
% 	\end{bmatrix} \nonumber
% $}
% \end{align}
% \begin{align}
% \scalebox{0.85}{
% $
% 	\bR_X(\gamma)=
% 	\begin{bmatrix}
% 		1 & 0          & 0           \\
% 		0 & \cos\gamma & -\sin\gamma \\
% 		0 & \sin\gamma & \cos\gamma
% 	\end{bmatrix} %\nonumber
% $}
% \end{align}

\begin{align}
	\bR_Z(\alpha)=
	\begin{bmatrix}
		\cos\alpha  & -\sin \alpha & 0 \\
		\sin \alpha & \cos\alpha   & 0 \\
		0           & 0            & 1
	\end{bmatrix}
\\
	\bR_Y(\beta)=
	\begin{bmatrix}
		\cos\beta  & 0 & \sin \beta \\
		0          & 1 & 0          \\
		-\sin\beta & 0 & \cos\beta
	\end{bmatrix}
\\
	\bR_X(\gamma)=
	\begin{bmatrix}
		1 & 0          & 0           \\
		0 & \cos\gamma & -\sin\gamma \\
		0 & \sin\gamma & \cos\gamma
	\end{bmatrix}
\end{align}

% The composite rotation matrix $\bR(\alpha,\beta,\gamma)$ is then $\bR(\alpha,\beta,\gamma)\triangleq\bR_Z(\alpha)\bR_Y(\beta)\bR_X(\gamma)$. The \ac{UE} orientation at time $t$ is determined by $\alpha_t$, $\beta_t$ and $\gamma_t$. We assume that the \ac{UE} has access to the erroneous estimates of these orientations, i.e., $\hat \alpha_t$, $\hat \beta_t$ and $\hat \gamma_t$. % use this for ieee

The composite rotation matrix $\bR(\alpha,\beta,\gamma)$ is then calculated as
\begin{align}
\bR(\alpha,\beta,\gamma)\triangleq\bR_Z(\alpha)\bR_Y(\beta)\bR_X(\gamma)
\end{align}
The \ac{UE} orientation at time $t$ is determined by $\alpha_t$, $\beta_t$ and $\gamma_t$. We assume that the \ac{UE} has access to the erroneous estimates of these orientations, i.e., $\hat \alpha_t$, $\hat \beta_t$ and $\hat \gamma_t$. % use this for arxiv

%%%%%%%%%%%%%%%%%%%%%%%%%%%%%%%%%%%%%%%%%%%%%%%%%%%%%%%%%%%%%%%%%%%%%%%%%%%%%%%%%%%%%%%%%%%%%%%%%%%%%%%%%%%%%
\section{Simulation setup} % page 3
\label{sec:simulation_setup}
%%%%%%%%%%%%%%%%%%%%%%%%%%%%%%%%%%%%%%%%%%%%%%%%%%%%%%%%%%%%%%%%%%%%%%%%%%%%%%%%%%%%%%%%%%%%%%%%%%%%%%%%%%

As the discussion of the data-driven approach merits the discussion of data-set preparation, we first provide the simulation setup, before discussing the data-driven approach. We provide only the most necessary simulation parameters in this article and refer the reader to~\cite{ali2021orientation} for the details.
	
\begin{figure}[t]
	\centering
    \includegraphics[width=0.45\textwidth]{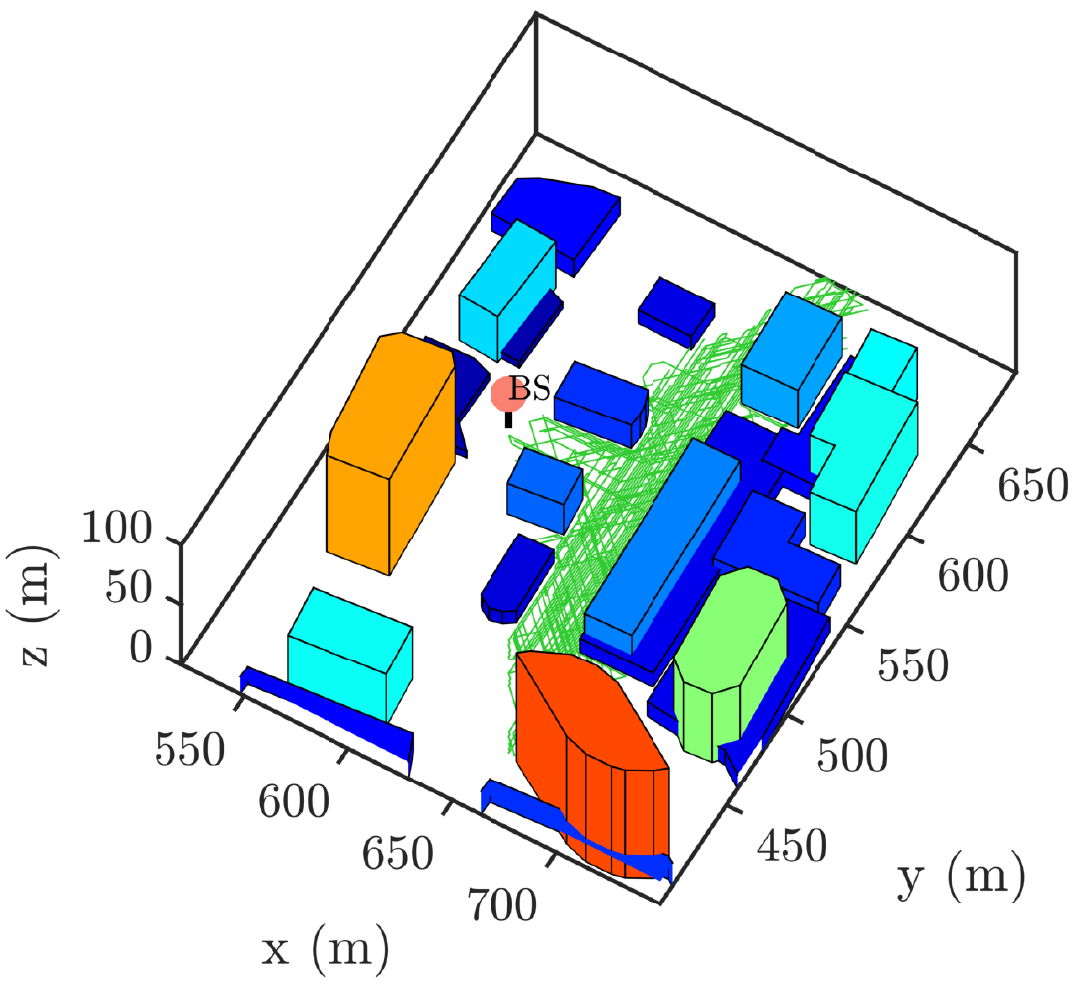} % for arxiv
	\caption{The ray-tracing simulation setup of downtown Rosslyn, VA. The location of \ac{BS} is shown with the salmon color disc, and the \ac{UE} moves along the trajectory shown by green lines.}
	\label{fig:WI_RT}
\end{figure}

The ray-tracing channels are generated for downtown Rosslyn, VA, USA, using Wireless InSite\textregistered~\cite{Remcom01} software. In Fig.~\ref{fig:WI_RT}, the \ac{BS} is shown with a salmon color disk. The \ac{UE} trajectory is shown via green colored lines. For generating the trajectory, the UE picks a random destination point in the cell, and once it reaches the picked point, it picks another random destination point. The A$^\ast$ search algorithm is used to find a short route from one randomly picked destination point to the next~\cite{hart1968formal}. The \ac{UE} trajectory is limited to a $120^{\circ}$ sector, and for the simulation, the \ac{UE} picks $200$ random destination points, and the total length of the trajectory is $20$ km.
	
\begin{figure*}[hbt!]  % use this figure for arxiv
	\centering
	\begin{subfigure}[t]{0.47\textwidth}
		\centering
		\captionsetup{width=.95\linewidth,font=footnotesize}
		\includegraphics[width=1\textwidth]{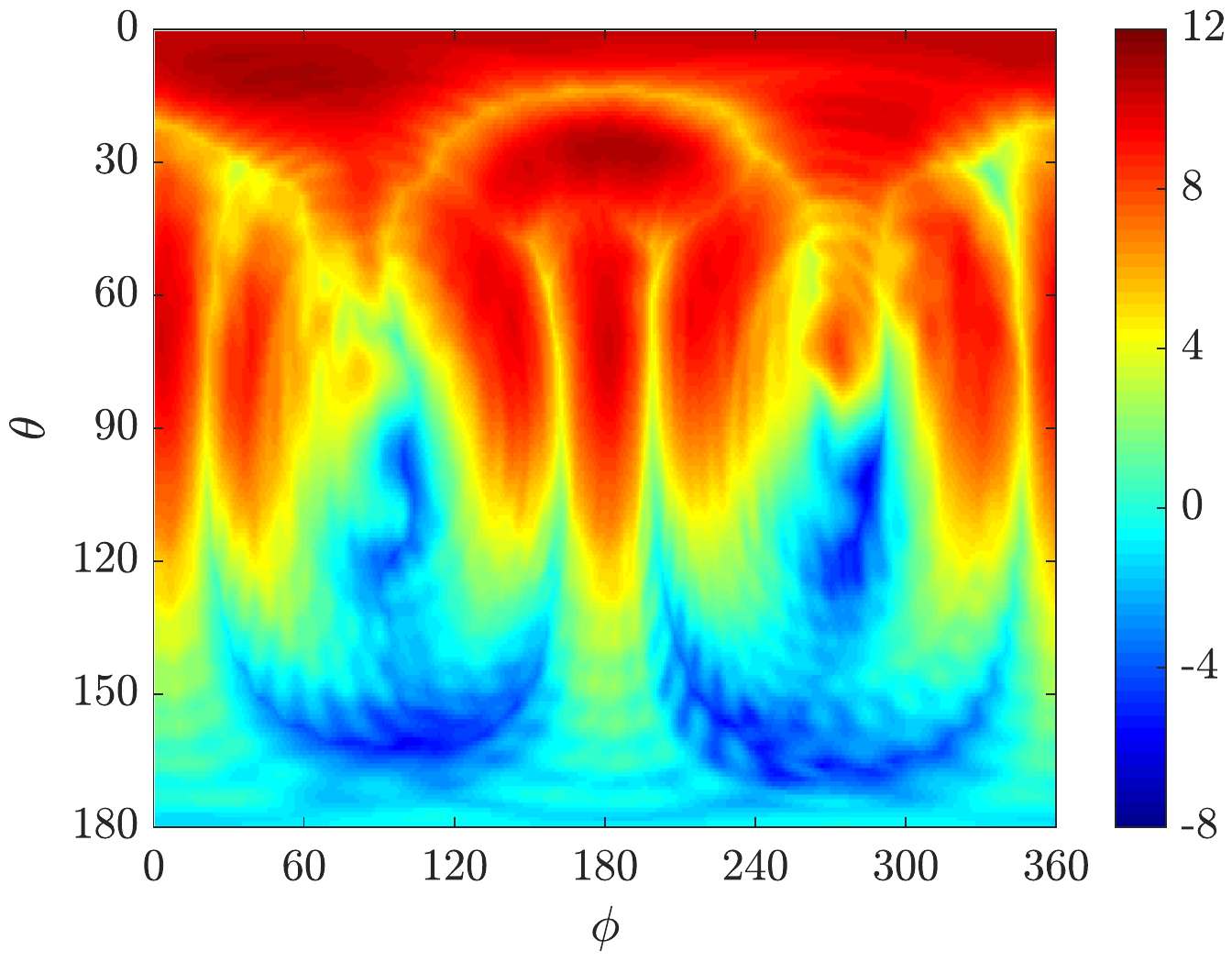}
		\caption{The composite radiation pattern of the \acp{WB}.}
		\label{fig:compWB}
	\end{subfigure}
	\begin{subfigure}[t]{0.47\textwidth}
		\centering
		\captionsetup{width=.95\linewidth,font=footnotesize}
		\includegraphics[width=1\textwidth]{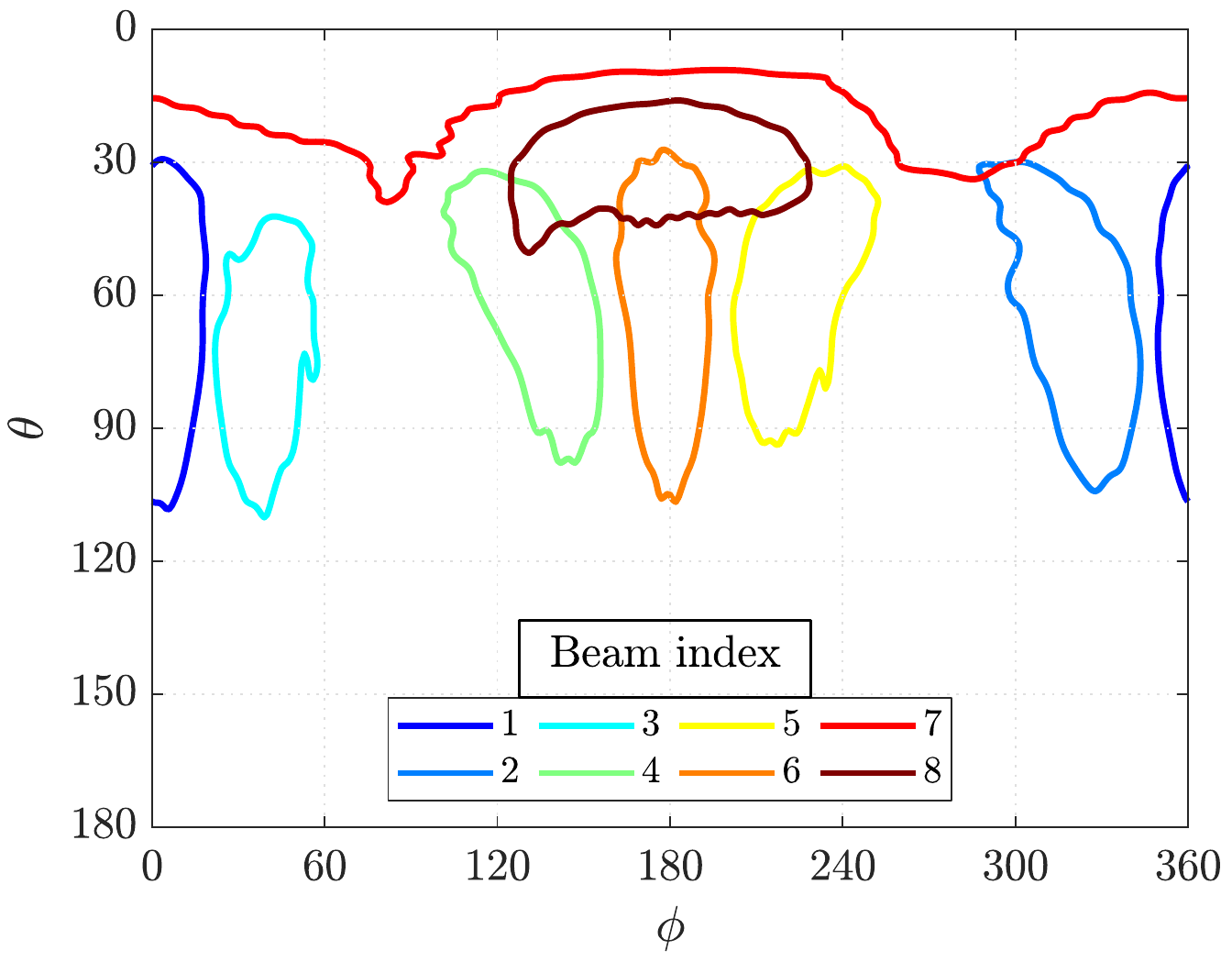}
		\caption{The $\SI{3}{\dB}$ contours of the \acp{WB}.}
		\label{fig:contWB}
	\end{subfigure}
	
	\begin{subfigure}[t]{0.47\textwidth}
		\centering
		\captionsetup{width=.95\linewidth,font=footnotesize}
		\includegraphics[width=1\textwidth]{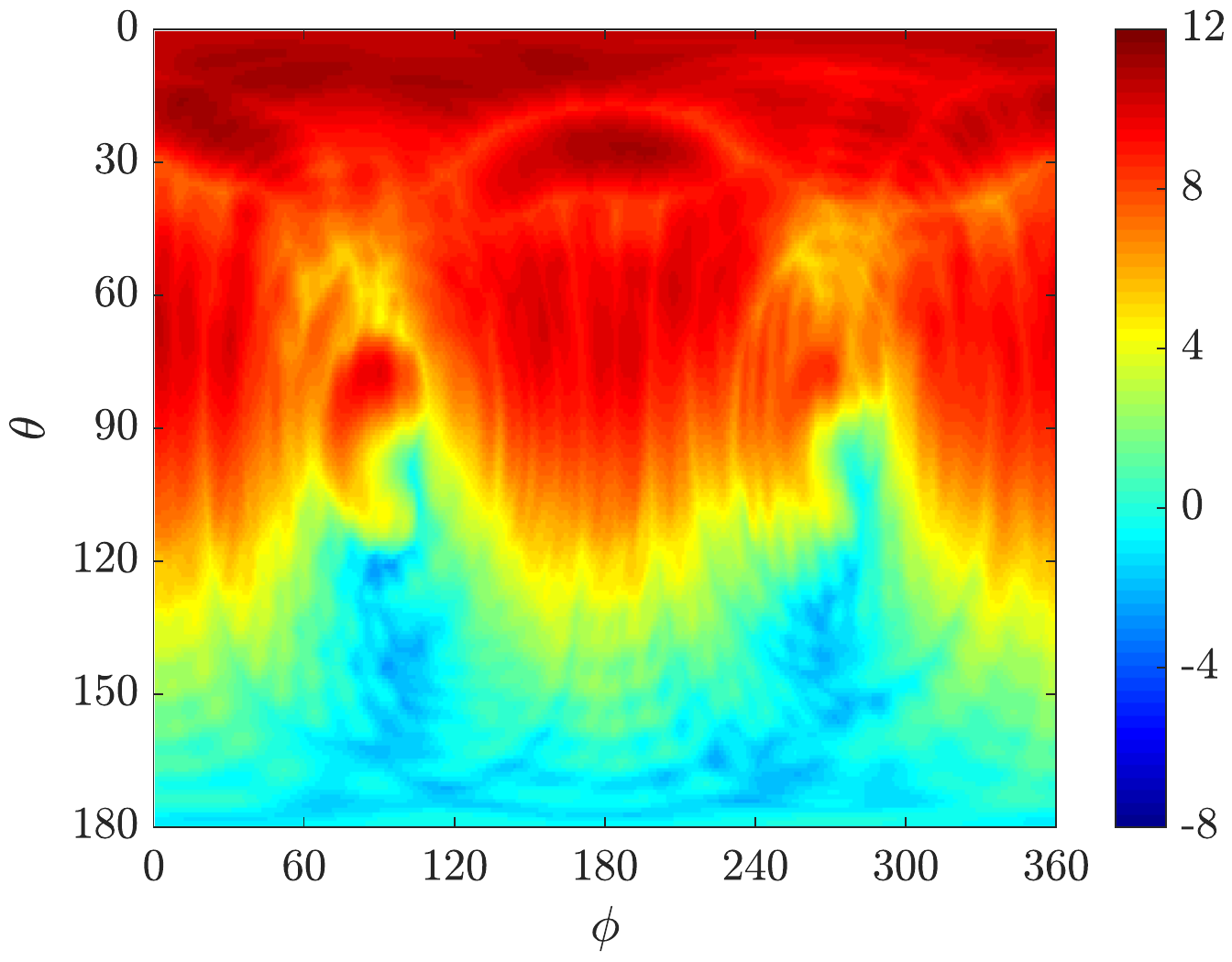}
		\caption{The composite radiation pattern of the \acp{NB}.}
		\label{fig:compNB}
	\end{subfigure}
	\begin{subfigure}[t]{0.47\textwidth}
		\centering
		\captionsetup{width=.95\linewidth,font=footnotesize}
		\includegraphics[width=1\textwidth]{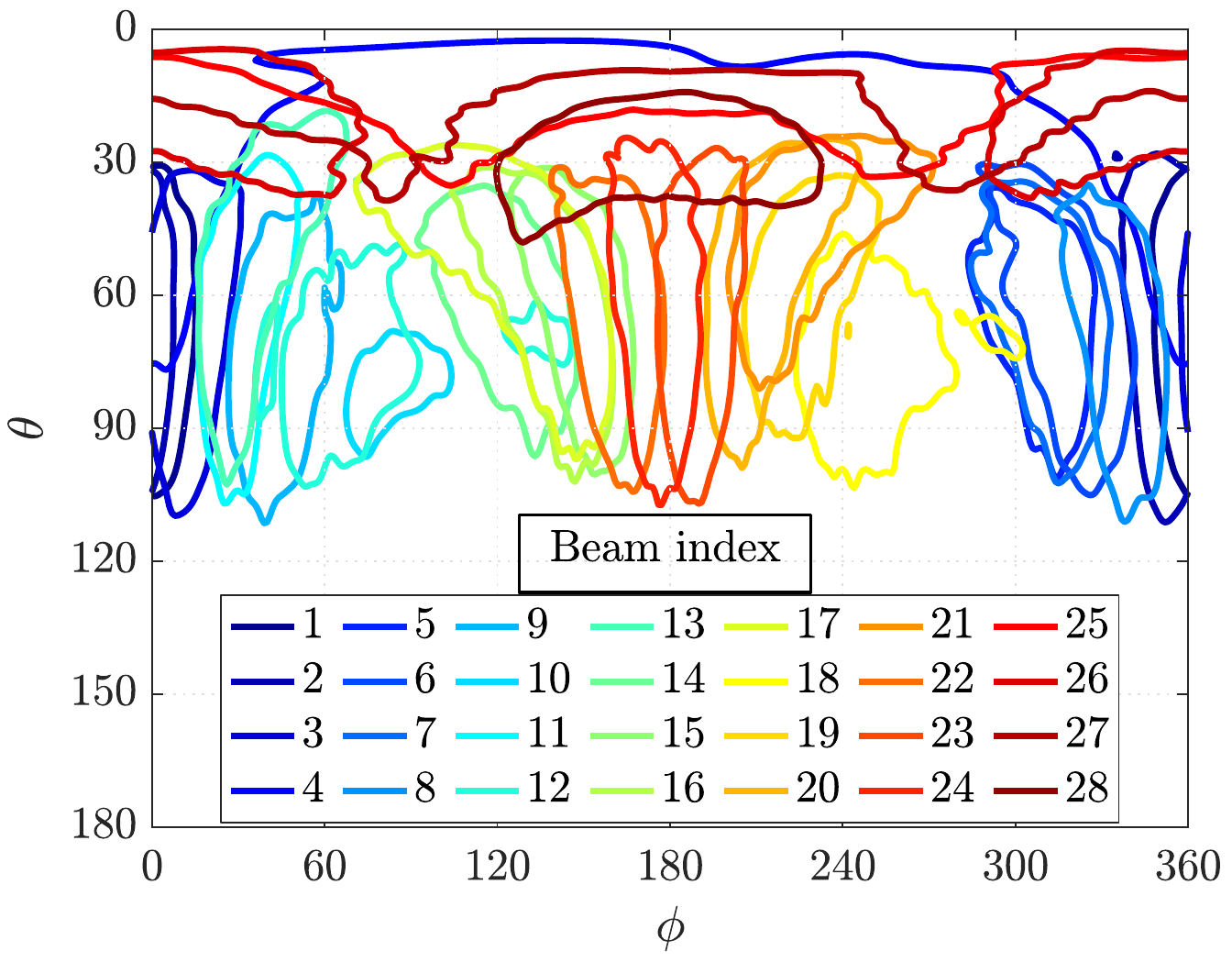}
		\caption{The $\SI{3}{\dB}$ contours of the \acp{NB}.}
		\label{fig:contNB}
	\end{subfigure}
	\hfill
	\caption{The composite radiation patterns $\SI{3}{\dB}$ contours of the $\MUE^{\rmW}=8$ \acp{WB} and $\MUE^{\rmN}=28$ \acp{NB}. The composite radiation pattern is plotted in the dB scale and the beam indices for the contour plots are given in the legend. They are based on a realistic phone setup with three $1\times 4$ ULA arrays on the left edge, right edge and the back of the phone, respectively.}
	\label{fig:patternsandcontours}
\end{figure*}

The operating frequency is~\SI{28}{\giga\hertz}, \ac{BW} is $\SI{100}{\mega\hertz}$, \ac{SCS} is $\SI{240}{\kilo\hertz}$, and the transmit power is $\PT=\SI{30}{\dBm}$. As our main focus is \ac{BM} at the \ac{UE} side, we assume a single isotropic antenna \ac{BS}, i.e., $\MBS=1$. The \ac{UE} either uses $\MUE^{\rmW}=8$ wide beams or $\MUE^{\rmN}=28$ narrow beams. The codebooks are obtained using the K-Means method of~\cite{mo2019beam}. The codebooks are designed assuming $\SI{3}{\bit}$ phase-shifters with no amplitude scaling. Fig.~\ref{fig:patternsandcontours} shows the $\SI{3}{\dB}$ contour plots of the wide and narrow beams. From beam measurements on a mobile device, it was observed that the measured \ac{RSRP} varies substantially over time, even in a static setup. We use the method discussed in~\cite{ali2021orientation} to generate the noise $n_t$ that models this variation. 
	
% \begin{figure}[t] % use this figure for ieee
% 	\centering
% % 	\begin{subfigure}[t]{0.24\textwidth}
% % 		\centering
% % 		\captionsetup{width=.95\linewidth,font=footnotesize}
% % 		\includegraphics[width=1\textwidth]{./Figures/CompositepatternWB.pdf}
% % 		\caption{The composite radiation pattern of the \acp{WB}.}
% % 		\label{fig:compWB}
% % 	\end{subfigure}
% 	\begin{subfigure}[t]{0.24\textwidth}
% 		\centering
% 		\captionsetup{width=.95\linewidth,font=footnotesize}
% 		\includegraphics[width=1\textwidth]{./Figures/ContourplotsWB.pdf}
% 		\caption{The $\SI{3}{\dB}$ contours of the \acp{WB}.}
% 		\label{fig:contWB}
% 	\end{subfigure}
% % 	\begin{subfigure}[t]{0.24\textwidth}
% % 		\centering
% % 		\captionsetup{width=.95\linewidth,font=footnotesize}
% % 		\includegraphics[width=1\textwidth]{./Figures/CompositepatternNB.pdf}
% % 		\caption{The composite radiation pattern of the \acp{NB}.}
% % 		\label{fig:compNB}
% % 	\end{subfigure}
% 	\begin{subfigure}[t]{0.24\textwidth}
% 		\centering
% 		\captionsetup{width=.95\linewidth,font=footnotesize}
% 		\includegraphics[width=1\textwidth]{./Figures/ContourplotsNB.pdf}
% 		\caption{The $\SI{3}{\dB}$ contours of the \acp{NB}.}
% 		\label{fig:contNB}
% 	\end{subfigure}
% 	\hfill
% 	\caption{The $\SI{3}{\dB}$ contours of the $\MUE^{\rmW}=8$ \acp{WB} and $\MUE^{\rmN}=28$ \acp{NB}. The beam indices for the contour plots are given in the legend. They are based on a realistic phone setup with three $1\times 4$ ULA arrays on the top, left, and right edge.}
% 	\label{fig:patternsandcontours}
% \end{figure}

We consider a filtered random walk based \ac{UE} orientation model. Specifically, consider a random walk $\bar\alpha_t=\bar\alpha_{t-1} + \cN(0,\sigma^2)$ where $\bar\alpha_0\sim\cU[0^{\circ},360^{\circ}]$. Then, the filtered random walk is $\alpha_t=\frac{1}{K}\sum_{k=0}^{K-1}\bar\alpha_{t-k}$
where $K$ is the filter length. A large value of $\sigma$ implies fast rotation, and a large $K$ implies smooth variation. The same procedure is followed for generating $\alpha_t$, $\beta_t$, and $\gamma_t$, and also a same value of $\sigma$ is used. 
%In experiments, $\sigma\in\{1^\circ,10^\circ \}$ and $K\in\{5,21\}$. 
We model the \ac{IMU} error as zero mean white Gaussian~\cite{kok2017using}. The level of error in each axis is then determined by the standard deviation, i.e., $\sigma_\alpha=2^\circ$, $\sigma_\beta=1^\circ$ and $\sigma_\gamma=1^\circ$.

We create four test cases shown in Table~\ref{tab:cases} to concretely capture the different levels of rotation speed, \ac{RSRP} information rates, and orientation smoothness. The rotation speed is either ``Slow'', i.e., $\sigma=1^\circ$ per $20$ ms, or ``Fast'', i.e., $\sigma=10^\circ$ per $20$ ms. The \ac{RSRP} information rate is either ``Normal'', i.e., $f=1$ and we get an \ac{RSRP} measurement every $\TSS$, or ``Sporadic'', i.e., $f=3$ and we get an \ac{RSRP} measurement every $3\TSS = 60$ ms. The rotation is either ``Smooth'', i.e., $K=21$, or ``Non-smooth'', i.e., $K=5$. Finally, a higher case index is a more favorable scenario for orientation-information use.
	
\begin{table}[t]
    \centering
	\caption{The four cases with slow or fast rotation speed, normal or sporadic \acp{RSRP} information, and smooth or non-smooth rotation.}
	\label{tab:cases}
	\resizebox{0.47\textwidth}{!}{%
		\begin{tabular}{|c|c|c|c|c}
			\cline{1-4}
			\textbf{Case} & \textbf{Rotation speed ($\sigma$)} & \textbf{RSRP information rate ($f$)} & \textbf{Rotation smoothness ($K$)} & \\ \cline{1-4}
			\textbf{1}    & Slow ($1^\circ$)                     & Normal (1)                             & Smooth (21)                          & \\ \cline{1-4}
			\textbf{2}    & Fast ($10^\circ$)                    & Normal (1)                             & Smooth (21)                          & \\ \cline{1-4}
			\textbf{3}    & Fast ($10^\circ$)                    & Sporadic (3)                           & Smooth (21)                          & \\ \cline{1-4}
			\textbf{4}    & Fast ($10^\circ$)                    & Sporadic (3)                           & Non-smooth (5)                       & \\ \cline{1-4}
		\end{tabular}%
	}
\end{table}
	
%%%%%%%%%%%%%%%%%%%%%%%%%%%%%%%%%%%%%%%%%%%%%%%%%%%%%%%%%%%%%%%%%%%%%%%%%%%%%%%%%%%%%%%%%%%%%%%%%%%%%%%%%%%%%
\section{Data-driven beam management} % page 3
\label{sec:dt_beam_management}
%%%%%%%%%%%%%%%%%%%%%%%%%%%%%%%%%%%%%%%%%%%%%%%%%%%%%%%%%%%%%%%%%%%%%%%%%%%%%%%%%%%%%%%%%%%%%%%%%%%%%%%%%%%%%
We divide the discussion of the data-driven \ac{BM} into two parts. First, we discuss the data preparation and partition, and then we discuss in detail the deep learning approach that is used for our \ac{BM} solution.
%-------------------------------------------------------------------------------------------
\subsection{Data preparation and partition}
%-------------------------------------------------------------------------------------------
In order to use both \ac{RSRP} and \ac{IMU} information, the data from these sensors need to be processed and fused together. For the \ac{RSRP} information, note that the \ac{UE} receives the \ac{SSB} by varying the receive beams in a round-robin manner\footnote{In this paper, we do not consider the scenario where the UE changes the receive beam sweeping order. However, as long as the same beam sweeping order is adopted in the training and operation phases, our RNN is applicable.}.
Specifically, the beam index of the beam used at time $t$ is $j_t=\bmod(\frac{t}{f},\MUE)+1 \in \{1,\cdots,\MUE\}$ when $t$ is an integer multiple of $f$ ($f=1$ or $3$ in the simulation), and the \ac{RSRP} measured on the beam $j_t$ is $s_{j_t}$. To prepare the data for the \ac{RNN}, an $\MUE\times 1$ vector is created with the $j_t$th entry set to $s_{j_t}$, and all the other entries set to $0$. For the normal measurement mode, a new \ac{RSRP} corresponding to an updated $j_t$ is available for every $t$, and the updated $1 \times \MUE$ vector is fed to the \ac{RNN}. For sporadic measurement mode, the $1 \times \MUE$ vector at each time step, no matter it is updated or not, is fed to the \ac{RNN}. For the \ac{IMU} information a new measurement is available every time step regardless of the \ac{RSRP} information rate. The \ac{IMU} information is captured directly through the matrix $\bR(\alpha,\beta,\gamma)$ discussed in Section~\ref{sec:IMUinfo}. This $3\times 3$ rotation matrix is flattened to get a $1\times 9$ vector. Therefore, after concatenation, the \ac{RSRP} information vector $1\times \MUE$ and rotation vector $1\times 9$, make a $1\times (\MUE+9)$ input vector for the \ac{RNN}. The output at each time step is the one hot encoded best beam index $\hat{j^\star}$.

\begin{table*}[htb!]
%	\vspace{0.3in} % added as edas wsa giving an error on top margin for page 6
\caption{The performance comparison of orientation-assisted \ac{BM} strategies (PF is Particle Filter and RNN is Recurrent Neural Network) in comparison with RSRP-only \ac{BM} in terms of beam prediction accuracy (\%), mean RSRP (dBm), and RSRP loss (dB). The performance is compared for two different \ac{UE} movement speeds and four cases outlined in Table~\ref{tab:cases}.}
\label{tab:RSRPonlyvsorientationassisted}
\resizebox{\textwidth}{!}{
	\begin{tabular}{|c|c|c||c|c|c||c|c|c||c|c|c||c|c|c|}
		\hline
		\multicolumn{3}{|c||}{\textbf{Case}}                                           & \multicolumn{3}{c||}{\textbf{1}}             & \multicolumn{3}{c||}{\textbf{2}}             & \multicolumn{3}{c||}{\textbf{3}}             & \multicolumn{3}{c|}{\textbf{4}}             \\ \hline
		\multicolumn{3}{|c||}{\textbf{Metric}}                                         & \textbf{AC} & \textbf{RSRP} & \textbf{Loss} & \textbf{AC} & \textbf{RSRP} & \textbf{Loss} & \textbf{AC} & \textbf{RSRP} & \textbf{Loss} & \textbf{AC} & \textbf{RSRP} & \textbf{Loss} \\ \hline \hline
		\multirow{6}{*}{$\SI{20}{\kilo\meter\per\hour}$}  & \multirow{3}{*}{\textbf{WB}} & \textbf{RSRP-only} & 90.16       & -101.9        & 0.17          & 75.60       & -102.57       & 0.83          & 49.10       & -104.91       & 3.17          & 40.43       & -105.90       & 4.15          \\ \cline{3-15} 
		&                              & \textbf{PF}        & 73.42       & -102.57       & 0.84          & 70.65       & -102.84       & 1.10          & 54.80       & -104.53       & 2.78          & 50.58       & -104.97       & 3.22          \\ \cline{3-15} 
		&                              & \textbf{RNN}       & 88.88       & -101.9        & 0.17          & 80.71       & -102.22       & 0.47          & 80.71       & -102.22       & 0.47          & 60.45       & -103.69       & 1.93          \\ \cline{2-15} 
		& \multirow{3}{*}{\textbf{NB}} & \textbf{RSRP-only} & 63.49       & -100.98       & 0.86          & 28.55       & -104.51       & 4.42          & 13.11       & -107.63       & 7.54          & 11.14       & -108.04       & 7.94          \\ \cline{3-15} 
		&                              & \textbf{PF}        & 37.14       & -102.99       & 2.87          & 33.05       & -103.38       & 3.29          & 21.72       & -105.56       & 5.47          & 21.41       & -105.66       & 5.57          \\ \cline{3-15} 
		&                              & \textbf{RNN}       & 57.52       & -101.02       & 0.90          & 43.99       & -101.98       & 1.88          & 30.08       & -103.51       & 3.42          & 26.69       & -103.99       & 3.90          \\ \hline \hline
		\multirow{6}{*}{$\SI{60}{\kilo\meter\per\hour}$}  & \multirow{3}{*}{\textbf{WB}} & \textbf{RSRP-only} & 85.36       & -102.59       & 0.41          & 72.20       & -103.06       & 1.09          & 44.16       & -105.65       & 3.68          & 36.87       & -106.53       & 4.56          \\ \cline{3-15} 
		&                              & \textbf{PF}        & 69.06       & -103.36       & 1.18          & 67.83       & -103.30       & 1.33          & 49.75       & -105.28       & 3.31          & 47.19       & -105.56       & 3.60          \\ \cline{3-15} 
		&                              & \textbf{RNN}       & 84.53       & -102.55       & 0.36          & 78.18       & -102.63       & 0.64          & 78.18       & -102.63       & 0.64          & 57.09       & -104.24       & 2.26          \\ \cline{2-15} 
		& \multirow{3}{*}{\textbf{NB}} & \textbf{RSRP-only} & 53.39       & -102.21       & 1.76          & 26.36       & -105.10       & 4.84          & 10.74       & -108.33       & 8.07          & 9.53        & -108.73       & 8.43          \\ \cline{3-15} 
		&                              & \textbf{PF}        & 31.62       & -103.93       & 3.48          & 30.31       & -103.95       & 3.69          & 18.73       & -106.46       & 6.20          & 18.34       & -106.50       & 6.20          \\ \cline{3-15} 
		&                              & \textbf{RNN}       & 51.58       & -101.91       & 1.44          & 41.35       & -102.47       & 2.20          & 28.03       & -104.10       & 3.84          & 25.40       & -104.53       & 4.22          \\ \hline \hline
		\multirow{6}{*}{$\SI{100}{\kilo\meter\per\hour}$} & \multirow{3}{*}{\textbf{WB}} & \textbf{RSRP-only} & 81.07       & -102.43       & 0.66          & 69.31       & -103.22       & 1.30          & 41.12       & -106.03       & 4.11          & 34.52       & -106.78       & 4.86          \\ \cline{3-15} 
		&                              & \textbf{PF}        & 67.46       & -103.12       & 1.35          & 65.03       & -103.52       & 1.60          & 45.29       & -105.76       & 3.84          & 42.98       & -106.02       & 4.10          \\ \cline{3-15} 
		&                              & \textbf{RNN}       & 80.65       & -102.34       & 0.54          & 74.36       & -102.80       & 0.86          & 74.36       & -102.80       & 0.86          & 52.33       & -104.68       & 2.74          \\ \cline{2-15} 
		& \multirow{3}{*}{\textbf{NB}} & \textbf{RSRP-only} & 46.73       & -102.66       & 2.47          & 23.86       & -105.60       & 5.26          & 9.88        & -108.52       & 8.17          & 9.45        & -108.86       & 8.48          \\ \cline{3-15} 
		&                              & \textbf{PF}        & 29.24       & -104.28       & 4.09          & 27.79       & -104.64       & 4.30          & 17.00       & -106.97       & 6.63          & 15.98       & -107.14       & 6.76          \\ \cline{3-15} 
		&                              & \textbf{RNN}       & 45.38       & -102.11       & 1.89          & 37.09       & -103.00       & 2.64          & 25.02       & -104.69       & 4.33          & 23.07       & -105.08       & 4.68          \\ \hline
	\end{tabular}
}
\end{table*}

Three \ac{UE} movement speeds were used, i.e., $\SI{20}{\kilo\meter\per\hour}$, $\SI{60}{\kilo\meter\per\hour}$, and $\SI{100}{\kilo\meter\per\hour}$. For each speed, we wanted to have around a million data points. As the number of sample points on the $\SI{20}{\kilo\meter}$ trajectory with sampling interval $20$ ms are around $180,000$ for \ac{UE} speed of $\SI{20}{\kilo\meter\per\hour}$, we used $6$ independent realizations of the orientation to get around a million points. Similarly for $\SI{60}{\kilo\meter\per\hour}$ and $\SI{100}{\kilo\meter\per\hour}$, we needed around $17$ and $28$ independent realizations of the orientation to get around a million points. Note that the UE rotation speed (either $1^\circ$ or $10^\circ$ per $20$ ms) is independent of the UE movement speed.

We partitioned the data across the trajectory. This implied that there was a high chance of having wireless channels in the testing set that were not seen by the \ac{RNN} through the training set. The training, validating, and test data split is around $70\%$, $20\%$, and $10\%$. Note that we have data for three different \ac{UE} speeds, two different rotation speeds, two different \ac{RSRP} information rates, and two levels of smoothness. We combine all this data to increase the training data size for \ac{RNN}.

%-------------------------------------------------------------------------------------------
\subsection{The deep learning approach}
\begin{figure}[t]
	\centering
	\includegraphics[width=0.47\textwidth]{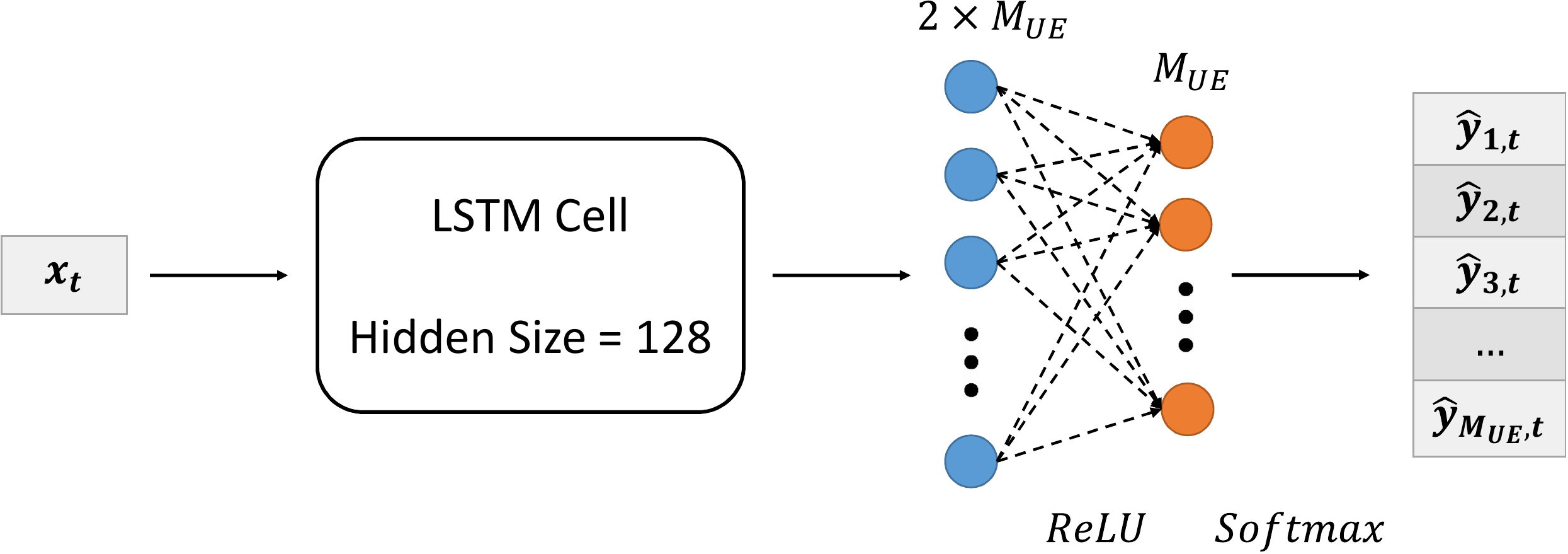}
	\caption{The \ac{RNN} architecture consists of a \ac{LSTM} cell and two \ac{FC} layers, where $\mathbf{x}_t$ (size: $1\times (\MUE+9)$) is the input and $\hat{\mathbf{y}}_t$ (size: $1\times \MUE$) is the beam prediction at time $t$.}
	\label{fig:NN}
\end{figure}

%-------------------------------------------------------------------------------------------
To handle the BM problem in a data-driven manner, we utilized supervised deep learning. We designed and used a \ac{RNN} architecture to perform beam prediction. Since this is a time-series classification problem, using \ac{RNN} is logical as \ac{RNN} exhibits temporal dynamic behavior which allows it to learn and process the temporal relation in the input sequences. Our \ac{RNN} architecture is shown in Fig.~\ref{fig:NN}. We used this architecture because it yielded the best results and it is relatively lightweight. The architecture includes a \ac{LSTM} cell \cite{hochreiter1997long} with a hidden size of $128$ neurons. This is followed by a \ac{FC} layer of size $2\times \MUE$ with a ReLU activation and another \ac{FC} layer of size $\MUE$ with a soft-max activation for multi-class classification purpose.

The \ac{RNN} adopts the categorical cross-entropy loss function. Specifically, the loss $L_t$ at each time step $t$ is calculated as 
% \begin{align}
% \scalebox{1}{
% $L_t=-\sum_{m=1}^{\MUE} y_{t,m}\log(\hat{y}_{t,m})$ %\nonumber
% }
% \end{align}
%
\begin{align}
L_t=-\sum_{m=1}^{\MUE} y_{t,m}\log(\hat{y}_{t,m}) %\nonumber
\end{align}
where $y_{t,m}$ is the corresponding target value at time step $t$ of class $m$, and $\hat{y}_{t,m}$ is the predicted probability of class $m$ at time $t$. Additionally, Adam optimizer %\cite{kingma2014adam} 
with a learning rate of $0.001$ was used. The training took $10,000$ epochs to converge with a batch size of $6$ trajectories (sequences).

\begin{figure}[t]
	\centering
	\includegraphics[width=0.47\textwidth]{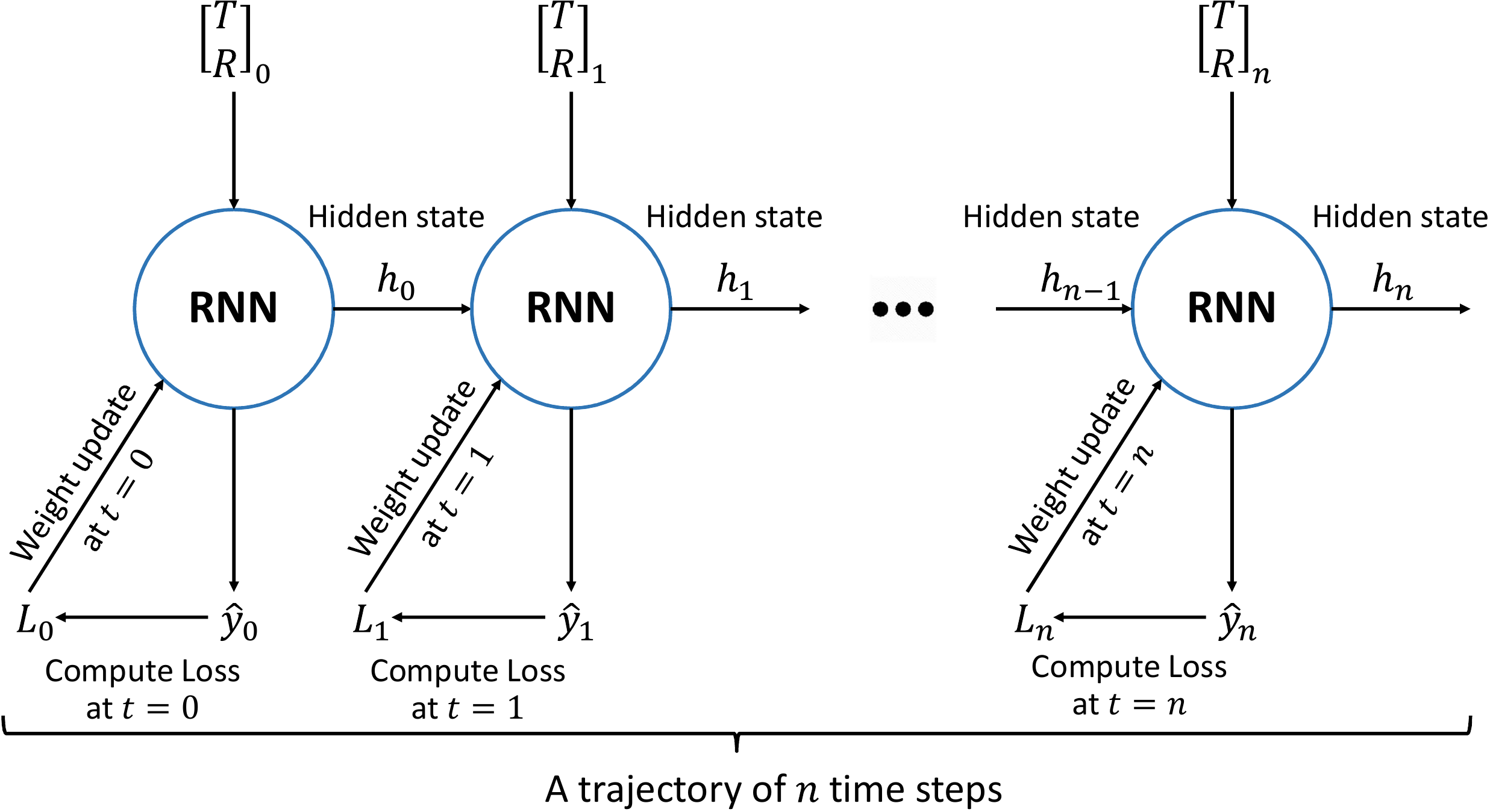}
	\caption{The training and inference step. For the training step, at each time step $t$, the input $[\mathbf{T}, \mathbf{R}]_t$ is fed into the \ac{RNN}. A beam decision $\hat{y}_t$ will be produced. The loss $L_t$ shall then be calculated and used to update the \ac{RNN}. For the inference step, the process stops at the beam decision.}
	\label{fig:inference}
\end{figure}

In the inference step, full trajectories (time sequences) were fed into the \ac{RNN}. At each time step $t$ within a trajectory, a beam decision will be produced. Fig. \ref{fig:inference} describes the training and inference process. The input at each step consist of 2 components. The first component is an array $\mathbf{T}_t$ of size $\MUE$ with the RSRP value of the last measured beam is set at that beam index $i$ and the values at other indices are set to 0. The second component is the rotation matrix computed from the current IMU orientation $\mathbf{R}_t$. The output at each time step is $\hat{y}_t$ which is the beam index of the best beam in the form of a one hot encoded vector. Additionally, the hidden state of the \ac{LSTM} cell $h_t$ at each time step is also returned to re-fed into the \ac{RNN}.

%%%%%%%%%%%%%%%%%%%%%%%%%%%%%%%%%%%%%%%%%%%%%%%%%%%%%%%%%%%%%%%%%%%%%%%%%%%%%%%%%%%%%%%%%%%%%%%%%%%%%%%%%%%%%
\section{Simulation results} % (page 4 and part of 5)
\label{sec:simulation_results}
%%%%%%%%%%%%%%%%%%%%%%%%%%%%%%%%%%%%%%%%%%%%%%%%%%%%%%%%%%%%%%%%%%%%%%%%%%%%%%%%%%%%%%%%%%%%%%%%%%%%%%%%%%%%%
We now present the simulation results that show the supremacy of the proposed data-driven strategy compared to conventional \ac{BM} and \ac{PF} based \ac{BM} strategy of~\cite{ali2021orientation}. For this purpose, we use three metrics as follows.
\begin{enumerate}
	\item Beam prediction accuracy or ``AC'' is the percentage of times the beam-predicted by a practical strategy, e.g., ``\ac{RSRP}-only'' or `Orientation-assisted'' is the same as the ``genie-aided'' best beam. The ``genie-aided'' best beam is obtained assuming instantaneous \ac{RSRP} knowledge of all the beams at the \ac{UE}
	\item The mean \ac{RSRP} or ``RSRP''.
	\item ``\ac{RSRP} loss'' or ``Loss'' is obtained by subtracting the \ac{RSRP} of a predicted beam from the \ac{RSRP} of the genie-aided best beam and taking the mean of the difference. 
\end{enumerate}

From the results in Table~\ref{tab:RSRPonlyvsorientationassisted}, we see that for Case $1$, the accuracy and the mean \ac{RSRP} of the orientation-assisted classical signal processing method, i.e., \ac{PF} strategy in~\cite{ali2021orientation}, is lower compared to \ac{RSRP}-only. This is because Case $1$ is not a favorable scenario for the use of orientation information. Specifically, the rotation speed itself is slow so there is not much benefit of using rotation information. In other words, before the mobile significantly rotates, \ac{RSRP} information on all (or most) beams can be collected to maintain relatively good \ac{BM}. Further, as the orientation information is erroneous, using orientation information can harm \ac{BM}. Even in this unfavorable case for the signal processing approach, the proposed machine learning method, i.e, deep learning strategy, can still work. Although the accuracy is a slightly lower than that of \ac{RSRP}-only, the mean \ac{RSRP}, which is a relevant metric for the communication, is very close or even better than \ac{RSRP}-only. 
	
In Case $2$, the \ac{RNN} performs better than the \ac{RSRP}-only, whereas the \ac{PF} does not. In Case $3$ and $4$ the \ac{RSRP} information rate is lower than the orientation information rate and the benefit of using the orientation-assisted strategies for \ac{BM} becomes clear. We can see that depending on the user movement speed and orientation smoothness, using orientation-assisted strategies can improve the beam prediction accuracy significantly. For example, at (Case 3, $\SI{60}{\kilo\meter\per\hour}$, \ac{WB}), the accuracy increases by $34\%$ from $44.16\%$ to $78.18\%$. 
The mean \ac{RSRP} is also boosted. For example, at (Case 4, $\SI{60}{\kilo\meter\per\hour}$, \ac{NB}), the mean \ac{RSRP} improves by $\SI{4.2}{\dB}$ from $\SI{-108.73}{\dBm}$ to $\SI{-104.53}{\dBm}$.

Finally, note that the performance of the proposed deep learning strategy is consistently better than the \ac{PF} strategy across all the scenarios. Moreover, the complexity of the deep learning strategy in terms of run-time cost is better in our experiments. The \ac{PF} strategy uses $1000$ particles to track the angle of arrival in the local coordinate system and updates their next values at each time step which might be problematic in terms of latency requirement. On the other hand, the deep learning strategy, although needs a large training overhead, is very efficient for online beam prediction. Our approach has the potential to be used in more general scenarios besides our experiment given that a larger and more diverse dataset is provided for the training purpose. Additionally, transfer learning \cite{tan2018survey} as well as meta learning \cite{chen2019closer} can also be used to quickly adjust the model to satisfy the requirement of the changing problem.

%%%%%%%%%%%%%%%%%%%%%%%%%%%%%%%%%%%%%%%%%%%%%%%%%%%%%%%%%%%%%%%%%%%%%%%%%%%%%%%%%%%%%%%%%%%%%%%%%%%%%%%%%%%%%
\section{conclusion} % page 5 and references
\label{sec:conclusion}
%%%%%%%%%%%%%%%%%%%%%%%%%%%%%%%%%%%%%%%%%%%%%%%%%%%%%%%%%%%%%%%%%%%%%%%%%%%%%%%%%%%%%%%%%%%%%%%%%%%%%%%%%%%%%
We proposed a data-driven \ac{BM} strategy that jointly utilizes the \ac{RSRP} and orientation information through an \ac{RNN}. Specifically, we formulated the \ac{BM} problem as a classification problem where one class stands for a beam. The proposed strategy outperforms the conventional \ac{BM} which relies only on the RSRP measurement. In particular, the data-driven strategy can improve \ac{BM} accuracy by $34\%$ and boost the mean \ac{RSRP} by $\SI{4.2}{\dB}$ in challenging environments of high mobility and fast rotation UE and sporadic RSRP measurement. The $\SI{4.2}{\dB}$ gain is significant at the \ac{UE} because it is equivalent to a cut of uplink transmission power by 62\%, which substantially improves the \ac{UE} battery life. Furthermore, when both RSRP measurement and orientation information are utilized, the data-driven strategy performs consistently better than the model-based PF strategy. Lastly, the high-complexity \ac{RNN} training is done offline, and the data-driven strategy is more efficient than PF for online \ac{BM}.

The training of the \ac{RNN} is done with the simulation data of a single deployment area in the downtown area. More training data from other deployment areas, for example, suburban, rural, can be obtained from simulation and used to train a more robust \ac{RNN} for different propagation environments. This is left for future work. Another future direction is to implement and evaluate the proposed strategy in a mobile device. The 5G mmWave devices may have a different number of mmWave antenna arrays and mount them in different locations, thus the WB and NB radiation patterns will be different. Considering such kind of difference between the simulation and real deployment, training of the \ac{RNN} with more simulation data and possibly field measurement data, would be needed.
%%%%%%%%%%%%%%%%%%%%%%%%%%%%%%%%%%%%%%%%%%%%%%%%%%%%%%%%%%%%%%%%%%%%%%%%%%%%%%%%%%%%%%%%%%%%%%%%%%%%%%%%%%%%%

% \section*{Acknowledgment}

% The preferred spelling of the word ``acknowledgment'' in America is without 
% an ``e'' after the ``g''. Avoid the stilted expression ``one of us (R. B. 
% G.) thanks $\ldots$''. Instead, try ``R. B. G. thanks$\ldots$''. Put sponsor 
% acknowledgments in the unnumbered footnote on the first page.

% \section*{References}

\bibliographystyle{IEEEtran}
\bibliography{Master_Bibliography}

\begin{thebibliography}{10}
\providecommand{\url}[1]{#1}
\csname url@samestyle\endcsname
\providecommand{\newblock}{\relax}
\providecommand{\bibinfo}[2]{#2}
\providecommand{\BIBentrySTDinterwordspacing}{\spaceskip=0pt\relax}
\providecommand{\BIBentryALTinterwordstretchfactor}{4}
\providecommand{\BIBentryALTinterwordspacing}{\spaceskip=\fontdimen2\font plus
\BIBentryALTinterwordstretchfactor\fontdimen3\font minus
  \fontdimen4\font\relax}
\providecommand{\BIBforeignlanguage}[2]{{%
\expandafter\ifx\csname l@#1\endcsname\relax
\typeout{** WARNING: IEEEtran.bst: No hyphenation pattern has been}%
\typeout{** loaded for the language `#1'. Using the pattern for}%
\typeout{** the default language instead.}%
\else
\language=\csname l@#1\endcsname
\fi
#2}}
\providecommand{\BIBdecl}{\relax}
\BIBdecl

\bibitem{rappaport2013millimeter}
T.~S. Rappaport, S.~Sun, R.~Mayzus, H.~Zhao, Y.~Azar, K.~Wang, G.~N. Wong,
  J.~K. Schulz, M.~Samimi, and F.~Gutierrez, ``Millimeter wave mobile
  communications for 5g cellular: It will work!'' \emph{IEEE access}, vol.~1,
  pp. 335--349, 2013.

\bibitem{giordani2020toward}
M.~Giordani, M.~Polese, M.~Mezzavilla, S.~Rangan, and M.~Zorzi, ``Toward 6g
  networks: Use cases and technologies,'' \emph{IEEE Communications Magazine},
  vol.~58, no.~3, pp. 55--61, 2020.

\bibitem{Bjornson_MWC19}
E.~Bjornson, L.~Van~der Perre, S.~Buzzi, and E.~G. Larsson, ``Massive {MIMO} in
  sub-6 {GHz and mmWave}: Physical, practical, and use-case differences,''
  \emph{IEEE Wireless Communications}, vol.~26, no.~2, pp. 100--108, 2019.

\bibitem{Giordani2019Tutorial}
M.~Giordani, M.~Polese, A.~Roy, D.~Castor, and M.~Zorzi, ``A tutorial on beam
  management for {3GPP} nr at mmwave frequencies,'' \emph{IEEE Communications
  Surveys Tutorials}, vol.~21, no.~1, pp. 173--196, 2019.

\bibitem{Li_Access20}
Y.-N.~R. Li, B.~Gao, X.~Zhang, and K.~Huang, ``Beam management in
  millimeter-wave communications for {5G} and beyond,'' \emph{IEEE Access},
  vol.~8, pp. 13\,282--13\,293, 2020.

\bibitem{Heng_COMM21}
Y.~Heng, J.~G. Andrews, J.~Mo, V.~Va, A.~Ali, B.~L. Ng, and J.~C. Zhang, ``Six
  key challenges for beam management in {5.5G and 6G} systems,'' \emph{IEEE
  Communications Magazine}, vol.~59, no.~7, pp. 74--79, 2021.

\bibitem{Va_TVT18}
V.~Va, J.~Choi, T.~Shimizu, G.~Bansal, and R.~W. Heath, ``Inverse multipath
  fingerprinting for millimeter wave {V2I} beam alignment,'' \emph{IEEE TVT},
  vol.~67, no.~5, pp. 4042--4058, 2018.

\bibitem{Maschietti_GC17}
F.~Maschietti, D.~Gesbert, P.~de~Kerret, and H.~Wymeersch, ``Robust
  location-aided beam alignment in millimeter wave massive {MIMO},'' in
  \emph{GLOBECOM 2017 - 2017 IEEE Global Communications Conference}, 2017, pp.
  1--6.

\bibitem{Heng_TCCN21}
Y.~Heng and J.~G. Andrews, ``Machine learning-assisted beam alignment for
  {mmWave} systems,'' \emph{IEEE TCCN}, vol.~7, no.~4, pp. 1142--1155, 2021.

\bibitem{Ali2018Millimeter}
A.~Ali, N.~Gonzalez-Prelcic, and R.~W. Heath, ``Millimeter wave beam-selection
  using out-of-band spatial information,'' \emph{IEEE-J-WCOM}, vol.~17, no.~2,
  pp. 1038--1052, 2017.

\bibitem{Alrabeiah_TCOM20}
M.~Alrabeiah and A.~Alkhateeb, ``Deep learning for mmwave beam and blockage
  prediction using sub-6 {GHz} channels,'' \emph{IEEE Transactions on
  Communications}, vol.~68, no.~9, pp. 5504--5518, 2020.

\bibitem{kaya2021deep}
A.~{\"O}. Kaya and H.~Viswanathan, ``Deep learning-based predictive beam
  management for {5G} mmwave systems,'' in \emph{2021 IEEE WCNC}.\hskip 1em
  plus 0.5em minus 0.4em\relax IEEE, 2021, pp. 1--7.

\bibitem{Lim_TCOM21}
S.~H. Lim, S.~Kim, B.~Shim, and J.~W. Choi, ``Deep learning-based beam tracking
  for millimeter-wave communications under mobility,'' \emph{IEEE Transactions
  on Communications}, vol.~69, no.~11, pp. 7458--7469, 2021.

\bibitem{Echigo_TVT21}
H.~Echigo, Y.~Cao, M.~Bouazizi, and T.~Ohtsuki, ``A deep learning-based low
  overhead beam selection in {mmWave} communications,'' \emph{IEEE TVT},
  vol.~70, no.~1, pp. 682--691, 2021.

\bibitem{Ma_Ke_TCOM21}
K.~Ma, D.~He, H.~Sun, Z.~Wang, and S.~Chen, ``Deep learning assisted calibrated
  beam training for millimeter-wave communication systems,'' \emph{IEEE
  Transactions on Communications}, vol.~69, no.~10, pp. 6706--6721, 2021.

\bibitem{shim2014application}
D.-S. Shim, C.-K. Yang, J.~H. Kim, J.~P. Han, and Y.~S. Cho, ``Application of
  motion sensors for beam-tracking of mobile stations in mmwave communication
  systems,'' \emph{Sensors}, vol.~14, no.~10, pp. 19\,622--19\,638, 2014.

\bibitem{qi2018three}
Z.~Qi and W.~Liu, ``Three-dimensional millimetre-wave beam tracking based on
  smart phone sensor measurements and direction of arrival/time of arrival
  estimation for 5g networks,'' \emph{IET Microwaves, Antennas \& Propagation},
  vol.~12, no.~3, pp. 271--279, 2018.

\bibitem{brambilla2019inertial}
M.~Brambilla, M.~Nicoli, S.~Savaresi, and U.~Spagnolini, ``Inertial sensor
  aided mmwave beam tracking to support cooperative autonomous driving,'' in
  \emph{2019 IEEE ICC Workshops)}.\hskip 1em plus 0.5em minus 0.4em\relax IEEE,
  2019, pp. 1--6.

\bibitem{Rezaie2020Location}
S.~Rezaie, C.~N. Manch{\'o}n, and E.~De~Carvalho, ``{Location-and
  Orientation-Aided Millimeter Wave Beam Selection Using Deep Learning},'' in
  \emph{2020 IEEE International Conference on Communications, ICC 2020}, 2020,
  pp. 1--6.

\bibitem{ali2021orientation}
A.~Ali, J.~Mo, B.~L. Ng, V.~Va, and J.~C. Zhang, ``Orientation-assisted beam
  management for beyond 5g systems,'' \emph{IEEE Access}, vol.~9, pp.
  51\,832--51\,846, 2021.

\bibitem{kunsch2013particle}
H.~R. K{\"u}nsch, ``Particle filters,'' \emph{Bernoulli}, vol.~19, no.~4, pp.
  1391--1403, 2013.

\bibitem{3GPP38901}
\BIBentryALTinterwordspacing
3GPP, ``{Study on channel model for frequencies from 0.5 to 100 GHz},'' {3rd
  Generation Partnership Project (3GPP)}, TR 38.901, Dec. 2017, version 14.3.0.
  [Online]. Available: \url{http://www.3gpp.org/DynaReport/38901.htm}
\BIBentrySTDinterwordspacing

\bibitem{Remcom01}
{Remcom Inc.}, \emph{Wireless InSite 3.3.5 Reference Manual}, November 2020.

\bibitem{hart1968formal}
P.~E. Hart, N.~J. Nilsson, and B.~Raphael, ``A formal basis for the heuristic
  determination of minimum cost paths,'' \emph{IEEE transactions on Systems
  Science and Cybernetics}, vol.~4, no.~2, pp. 100--107, 1968.

\bibitem{mo2019beam}
J.~Mo, B.~L. Ng, S.~Chang, P.~Huang, M.~N. Kulkarni, A.~Alammouri, J.~C. Zhang,
  J.~Lee, and W.-J. Choi, ``Beam codebook design for 5g mmwave terminals,''
  \emph{IEEE Access}, vol.~7, pp. 98\,387--98\,404, 2019.

\bibitem{kok2017using}
M.~Kok, J.~D. Hol, and T.~B. Sch{\"o}n, ``Using inertial sensors for position
  and orientation estimation,'' \emph{arXiv preprint arXiv:1704.06053}, 2017.

\bibitem{hochreiter1997long}
S.~Hochreiter and J.~Schmidhuber, ``Long short-term memory,'' \emph{Neural
  computation}, vol.~9, no.~8, pp. 1735--1780, 1997.

\bibitem{tan2018survey}
C.~Tan, F.~Sun, T.~Kong, W.~Zhang, C.~Yang, and C.~Liu, ``A survey on deep
  transfer learning,'' in \emph{ICANN}.\hskip 1em plus 0.5em minus 0.4em\relax
  Springer, 2018, pp. 270--279.

\bibitem{chen2019closer}
W.-Y. Chen, Y.-C. Liu, Z.~Kira, Y.-C.~F. Wang, and J.-B. Huang, ``A closer look
  at few-shot classification,'' \emph{arXiv preprint arXiv:1904.04232}, 2019.

\end{thebibliography}

\end{document}